\documentclass[a4paper,11pt]{article}
\pdfoutput=1 

\usepackage{jheppub} 
\usepackage{slashed}

\usepackage[T1]{fontenc} 
\usepackage[normalem]{ulem}
\usepackage{braket}

\title{\boldmath Krylov complexity for Lin-Maldacena geometries and their holographic duals}

\author[a]{Dibakar Roychowdhury}

\affiliation[a]{Department of Physics, Indian Institute of Technology Roorkee,\\Roorkee 247667, Uttarakhand, India}

\abstract{We compute the rate of growth of operator size in matrix models by probing the Lin-Maldacena class of geometries with classical probes. We consider massive point particle probes whose proper momentum equals the size of the gauge invariant operator in the matrix model. We work out the example of the BMN Plane Wave Matrix Model using the electrostatic approach and the method of background fluxes. We also work out complexities in the D2 brane as well as NS5 brane limits of the BMN matrix model along with an example of the irrelevant deformation namely the non-Abelian T-dual of $AdS_5 \times S^5$. Finally, we carry out a possible calculation of the Krylov complexity on the matrix model counterpart by using a simple reduction ansatz known as the pulsating fuzzy sphere model. We outline an algorithm to define Krylov basis elements for the matrix model and compute a few Lanczos coefficients. Our analysis reveals that both the Krylov basis states as well as Lanczos coefficients are uniquely fixed in terms of the mass parameter of the matrix model.}
\begin{document} 
\maketitle
\flushbottom
\section{Introduction and General Idea}
There have been several proposals to compute holographic complexity in the literature, for example ``Complexity $=$ Volume'' \cite{Stanford:2014jda}, ``Complexity $=$ Action'' \cite{Brown:2015bva} and ``Complexity $=$ Anything'' \cite{Belin:2021bga}. The latest has been the proposal due to \cite{Caputa:2024sux}, which allows us to compute the Krylov complexity \cite{Parker:2018yvk}-\cite{Caputa:2021sib} in many body quantum mechanics by means of a dual gravitational theory. See \cite{Baiguera:2025dkc}-\cite{Nandy:2024evd}, for a nice set of comprehensive reviews on the subject.

The idea of this paper is based on a series of papers \cite{Susskind:2018tei}-\cite{Ageev:2018msv}, which argue about the size /momentum correspondence that eventually has been linked to complexity in the dual gauge theory. These ideas are built upon the fact that the proper momentum of a massive (particle) probe in the bulk equals the rate of growth of complexity (or the size) of the dual operator on the quantum mechanical counterpart, which have attracted renewed attention in recent years, particularly in the context of Holographic Krylov complexity growth of operators; see, for example, \cite{Fatemiabhari:2025poq}-\cite{Nastase:2026lhz}. There has been a parallel development to compute the holographic complexity for states; see, for example, \cite{Heller:2024ldz}-\cite{Ambrosini:2024sre}.

We extend the above ideas and perform a computation of Krylov operator complexity \cite{Caputa:2024sux} for BMN Plane Wave Matrix Models (PWMM) \cite{Berenstein:2002jq} and its irrelevant deformations \cite{Lozano:2017ole} in a holographic setup. We also discuss a field theory counterpart and a possible interpretation of our calculations in a matrix model framework. In the dual gravity picture, we identify the \emph{proper} radial coordinate ($\rho$) in the ($\sigma, \eta$) plane \cite{Lozano:2017ole} that serves as the basis for the Krylov operator complexity. We calculate the proper radial momentum ($\mathcal{P}_\rho$), which equals the rate of growth of the Krylov complexity \cite{Caputa:2024sux} in the dual matrix model. 

BMN matrix model \cite{Berenstein:2002jq} was originally conceived as a realization of M theory on a supersymmetric pp wave background in eleven dimensions \cite{Dasgupta:2002hx}, which turns out to be massive deformation of BFSS matrix model \cite{Banks:1996vh}. The above model has also been derived using the matrix regularization of the super-membrane on the pp wave \cite{Sugiyama:2002rs}. The gravity dual of the BMN matrix model (in type IIA) has been realized through a series of seminal works \cite{Lin:2004nb}-\cite{Lin:2004kw}, which was further extended in other directions \cite{Ling:2006up}-\cite{Amore:2024ihm}. 

In Section \ref{section 2}, we begin by considering the electrostatic description of the BMN matrix model, which is characterized by a set of conducting disks located along the holographic ($\eta$) axis, where ($\sigma , \eta$) serves as the electrostatic coordinates. This class of type IIA solutions is characterized by a potential function $V(\sigma , \eta)$ that satisfies Laplace's equation of electrostatics \cite{Lin:2005nh}, \cite{Nunez:2019gbg}. We probe this geometry with a massive particle that is dual to some local unitary operator $\mathcal{O}$ in the Large $N$ matrix model. The geometry in the UV approaches the metric of $N$ D0 branes, which reveals a linear rate of growth of complexity to begin with. As a special limit, we consider the D2 and NS5 brane limits \cite{Lin:2005nh} of the matrix matrix model. Although the complexity for the D2 brane solution approaches saturation at late times, it seems to increase at late times in the NS5 limit.

We revisit our calculations in Section \ref{section 3}, following Lin's approach \cite{Lin:2004kw} to the BMN mantrix model, which is based on the method of background fluxes and their fluctuations \cite{Polchinski:2000uf}. In particular, we explore the late time behavior of the particle geodesic near the shell of D2 branes. As the particle approaches the shell of the D2 branes, the corresponding rate of growth of the complexity scales differently ($\sim t^{35/24}$) than its UV counterpart. The above scaling reveals a non-linear rate of growth of complexity as the particle approaches deep inside the bulk. This results from the fact that the geometry (in the interior of the bulk) is significantly modified as a result of the concentric shell of D2 branes. 

We repeat our calculations for the non-Abelian T-dual of $AdS_5 \times S^5$ in Section \ref{section 4}. This solution is dual to irrelevant deformation of the BMN matrix model, which corresponds to a smeared D0 brane solution in UV \cite{Lozano:2017ole}. The gravitational calculation reveals an increasing complexity for the dual matrix model at late times.

Finally, in Section \ref{section 5}, we bridge our gravity calculations with its matrix model counterpart by explicitly performing a calculation for the BMN matrix model using a simplified reduction ansatz, known as the pulsating fuzzy sphere model \cite{Asano:2015eha}-\cite{Amore:2024ihm}. We identify the corresponding Krylov basis for the operator complexity, which tri-diagonalizes the Liouvillian operator. We calculate the early time behavior of the Krylov complexity, which qualitatively agrees with gravity calculations. Our analysis reveals that the Lanczos coefficients (and hence the Krylov complexity) are fixed by the mass deformation parameter ($\mu$) of the matrix model. In particular, we compute the first two non zero coefficients $b_1$ and $b_2$. While $b_1 \propto \mu$ for the entire range of deformations, the other coefficient $b_2$ exhibits a nonlinear relationship with $\mu$, for generic values of the deformation parameter.

Before we proceed further, some important clarifications are in order.

$\bullet$ The geometry in the UV approaches the metric of $N$ D0 branes and is dual to a non-conformal theory. The theory has a mass gap characterized by a mass deformation parameter $\mu$, which corresponds to a massive deformation of the BFSS matrix model \cite{Banks:1996vh}. This is different from the previous analysis, where the dual field theory is conformal \cite{Caputa:2024sux}. However, as our analysis reveals, we have a universal feature, namely the complexity grows quadratically with time. This has been further confirmed through our toy model calculation in Section \ref{section 5}, see, for example, eq. \eqref{e5.44}. Also, we notice that the leading coefficient on the gravity side depends on the dipole deformation $P$ \eqref{e2.43}, while on the matrix model counterpart it corresponds to the massive deformation parameter $\mu$, see, for example, \eqref{e5.47}.

$\bullet$ One of the key ingredients in the gravity side of the calculation is to identify the (proper) radial coordinate. This is done by identifying the radial direction $r=\sqrt{\sigma^2 +\eta^2}$ in $(\sigma ,\eta)$ plane, where $(\sigma ,\eta)$ are the electrostatic coordinates. In the asymptotic limit ($r \rightarrow \infty$), these electrostatic coordinates ($ \sigma ,\eta$) combine nicely to produce the radial axis ($r$) associated with the near horizon geometry created by $N$ D0 branes \cite{Lozano:2017ole}, which serves as a basic motivation for us to consider particle geodesic in the ($ \sigma ,\eta$) plane.

$\bullet$ Finally, it is important to reemphasize that in the present paper we explore the Krylov complexity associated with operator growth $\mathcal{O}(t)=e^{iHt}\mathcal{O}(0)e^{-i H t}$ in quantum mechanics, along the lines of \cite{Parker:2018yvk},\cite{Hashimoto:2023swv}. The particle in the bulk corresponds to a local unitary operator $\mathcal{O}_0=\mathcal{O}(0)$ inserted at $t=0$ in the matrix model. In the Krylov approach towards complexity, we map this operator to the (initial) state $\mathcal{O}_0 \rightarrow\mathcal{O}_0\ket{vac}:=|\mathcal{O}_0)$ in the ``operator Hilbert space'' \cite{Parker:2018yvk}. Subsequently, the other states in the Hilbert space are constructed by acting on the commutator of the Hamiltonian $\hat{\mathcal{L}}^n \mathcal{O}_0 \ket{vac}:=|\mathcal{O}_n)$, where $\hat{\mathcal{L}}=[\hat{H},]$ is the Liouvillian operator. Here, $\ket{vac}$ is the vacuum (or the reference state) of the conventional Hilbert space, and $\hat{H}$ is the corresponding Hamiltonian. By Krylov complexity, we refer to the spread complexity of states $|\mathcal{O}_n)$ in the operator Hilbert space.
\section{PWMM and Krylov complexity in the Electrostatic approach}
\label{section 2}
We begin with a brief review of the field theory content. The matrix model has $SO(6)\times SO(3)$ global symmetry that preserves 16 (or $\mathcal{N}=2$) SUSYs. It is a $U(N)$ gauge theory of $N \times N$ matrices that allows several vacua (fuzzy spheres) that are in one to one correspondence with the partition of $N = \sum_k n_k N_k$, where $n_k$ is the multiplicity of the $k$ th irreducible representation of $SU(2)\sim SO(3)$ that has rank $N_k$. 

In the dual gravitational counter part, the role of $n_k$ is played by the charge on the D2 brane and the role of $N_k$ is set by the conducting disks located at discrete positions ($\eta_k\sim N_k$) along the $\eta$- axis. These conducting disks are characterized by NS5 branes that have a radius $R_k$ and a charge $Q_k \sim n_k$. 

The complete Type IIA solution that preserves $\mathcal{N}=2$ SUSY can be characterized by a potential function $V(\sigma ,\eta)$ that satisfies a Laplace equation
\begin{align}
    \ddot{V}+\sigma^2 V''=0
\end{align}
where we denote $\dot{V}=\sigma \partial_\sigma V$ and $V'=\partial_\eta V$.

For our purposes, we would be interested in the metric and the dilaton in the string frame, which can be expressed as \cite{Lin:2004nb}-\cite{Lin:2004kw}
\begin{align}
\label{e2.2}
    &ds_{string}^2=-f_1(\sigma,\eta)dt^2+f_2(\sigma , \eta)(d\sigma^2 + d \eta^2)+f_3 (\sigma , \eta)d\Omega_2^2+f_4(\sigma , \eta)d\Omega^2_5\\
    &e^\phi = f_5 (\sigma , \eta).
\end{align}

The individual functions are given by
\begin{align}
\label{e2.4}
    &f_1 (\sigma , \eta)=\frac{4 \ddot{V}}{\sqrt{-V''}\sqrt{\ddot{V}-2\dot{V}}}~;~f_2(\sigma , \eta)=\frac{2\sqrt{-V''}}{\dot{V}}\sqrt{\ddot{V}-2\dot{V}}\\
    &f_3(\sigma , \eta)=\frac{2\dot{V}V''}{\Delta}\frac{\sqrt{\ddot{V}-2\dot{V}}}{\sqrt{-V''}}~;~f_4(\sigma , \eta)=4\frac{\sqrt{\ddot{V}-2\dot{V}}}{\sqrt{-V''}}\\
    &f_5(\sigma , \eta)=\Big[ \frac{4(2 \dot{V}-\ddot{V})^3}{V'' \dot{V}^2 \Delta^2}\Big]^{1/4}~;~\Delta=(\ddot{V}-2\dot{V})V''-(\dot{V}')^2.
\end{align}

The potential function that asymptotes to metric of D0 brane is given by \cite{Lin:2005nh}
\begin{align}
\label{e2.7}
    V_{PWMM}(\sigma , \eta)=V_0 \Big(\eta \sigma^2-\frac{2}{3}\eta^3 \Big)+\frac{P \eta}{(\eta^2+\sigma^2)^{3/2}}=V_{D0}+\frac{P \eta}{(\eta^2+\sigma^2)^{3/2}}.
\end{align}

Given the potential function \eqref{e2.7}, the geometry \eqref{e2.2} in the UV ($\sigma , \eta \rightarrow \infty$) asymptotes to the near horizon geometry of $N\sim P \gg 1$ coincident $D0$ branes, which has a UV completion in terms of membranes (M2 or M5 branes) in 11d plane wave supergravity background in M-theory. Here $P\sim \sum_k \eta_k Q_k$ is the dipole moment produced by the conducting disks located at discrete locations $\eta_k$ and carrying a charge $Q_k$.
\subsection{Point particle dynamics}
The point particle motion is explored in the Einstein frame metric, which is given by
\begin{align}
\label{e2.8}
    ds^2_E=-h_1(\sigma , \eta)dt^2+h_2(\sigma, \eta)(d\sigma^2 + d \eta^2)+h_3 (\sigma , \eta)d\Omega_2^2+h_4(\sigma , \eta)d\Omega^2_5
\end{align}
where we denote the metric functions as $h_i(\sigma , \eta)=e^{-\phi/2}f_i(\sigma , \eta)$.

By holographic Krylov operator complexity, we refer to the point particle dynamics along the ($\sigma,\eta$) plane in the bulk description, where we introduce a proper momentum $\mathcal{P}_\rho = \mathcal{P}_\rho (\sigma , \eta)$ that equals the rate of growth of complexity $\dot{\mathcal{C}}(t)=\partial_t \mathcal{C}$ in the matrix model. The geodesic of the point particle is parametrized by the following choice of coordinates
\begin{align}
\eta =\eta (t)~;~\sigma = \sigma (t)
\end{align}
where all the remaining coordinates of the internal space are set to zero.

The metric induced on the particle world-line is given by
\begin{align}
    ds^2_{induced}=-\Big[ h_1(\sigma , \eta)-h_2(\sigma , \eta)(\dot{\sigma}^2+\dot{\eta}^2)\Big]dt^2=-h^{(ind)}_{tt}dt^2
\end{align}
where by the dot we mean derivative with respect to time.

The action of the point particle can be expressed as (we set $m=1$)
\begin{align}
\label{e2.11}
    &S_p=-\int dt \sqrt{h^{(ind)}_{tt}}=\int dt L\\
    &L=-\sqrt{h_1(\sigma , \eta)-h_2(\sigma , \eta)(\dot{\sigma}^2+\dot{\eta}^2)}.
    \label{e2.12}
\end{align}

The canonical momenta are given by\footnote{Notice that as the particle falls towards the interior of the bulk, therefore both $\dot{\sigma}<0$ and $\dot{\eta}<0$ are decreasing functions of time, which makes the momenta \eqref{e2.13}-\eqref{e2.14} negative. The negative sign here indicates that the momenta increase in the decreasing direction of the corresponding coordinate.}
\begin{align}
\label{e2.13}
    &P_\sigma = \frac{\partial L}{\partial \dot{\sigma}}=\frac{h_2(\sigma , \eta)\dot{\sigma}}{|L|}\\
    &P_\eta = \frac{\partial L}{\partial \dot{\eta}}=\frac{h_2(\sigma , \eta)\dot{\eta}}{|L|}.
    \label{e2.14}
\end{align}

The conserved Hamiltonian of the particle is given by
\begin{align}
\label{e2.15}
    H_0=P_\sigma \dot{\sigma}+P_\eta \dot{\eta}-L=\frac{h_1(\sigma , \eta)}{|L|}
\end{align}
which is a positive definite entity.

The equation of motion that follows from \eqref{e2.12}, can be expressed as
\begin{align}
\label{e2.16}
    \frac{d}{dt}\Big[ \frac{h_2\dot{\sigma}}{L}\Big]+\frac{1}{2L}\Big[ \partial_\sigma h_1 -\partial_\sigma h_2 (\dot{\sigma}^2+\dot{\eta}^2)\Big]=0\\
    \frac{d}{dt}\Big[ \frac{h_2\dot{\eta}}{L}\Big]+\frac{1}{2L}\Big[ \partial_\eta h_1 -\partial_\eta h_2 (\dot{\sigma}^2+\dot{\eta}^2)\Big]=0.
    \label{e2.17}
\end{align}
\subsection{Proper momentum}
The Krylov complexity \cite{Caputa:2024sux} in the bulk gravity dual can be obtained by noting the proper momentum ($\mathcal{P}_\rho$) along the geodesic of the particle. The proper momentum \cite{Fatemiabhari:2025poq}-\cite{Fatemiabhari:2025cyy} is associated with the proper distance $ds=d\rho$ along the massive particle trajectory which is obtained by setting $dt=0$ and all the directions associated with $S^2$ and $S^5$, which yields
\begin{align}
\label{e2.18}
    d\rho = \sqrt{h_2(\sigma , \eta)}\sqrt{d\sigma^2 + d\eta^2}.
\end{align}

Using \eqref{e2.18}, the proper momentum could be obtained as
\begin{align}
    \mathcal{P}_\rho = \frac{\partial L}{\partial \dot{\rho}}=P_\sigma \frac{\partial \dot{\sigma}}{\partial \dot{\rho}}+P_\eta \frac{\partial \dot{\eta}}{\partial \dot{\rho}}
\end{align}
which after a simplification reveals
\begin{align}
\label{e2.20}
    \mathcal{P}_\rho = \frac{\sqrt{h_2(\sigma , \eta)}}{|L|}\sqrt{\dot{\sigma}^2+\dot{\eta}^2}.
\end{align}

To obtain the proper momentum \eqref{e2.20}, one has to solve the equations of motion \eqref{e2.16}-\eqref{e2.17}, which is subject to the Hamiltonian constraint \eqref{e2.15}. Before we proceed further, it is customary to evaluate the metric functions $h_1$ and $h_2$ in the asymptotic limits. In order to do so, we introduce a new radial coordinate $r=\sqrt{\sigma^2 +\eta^2}$, which goes to infinity in the asymptotic limit $(\sigma , \eta) \rightarrow \infty$. A straightforward calculation reveals
\begin{align}
\label{e2.21}
   & h_1(r)|_{r \rightarrow \infty}=a_1 \frac{V_0^{11/8}}{P^{7/8}}r^{49/8}~;~a_1=\frac{8.2^{3/4}}{15^{7/8}}\\
   &h_2(r)|_{r \rightarrow \infty}=a_2 \frac{P^{1/8}V_0^{3/8}}{r^{7/8}}~;~a_2=2^{7/4} \sqrt[8]{15}
   \label{e2.22}
\end{align}
where $a_1$ and $a_2$ are some numerical pre-factors.

We solve the dynamical equations \eqref{e2.16}-\eqref{e2.17} considering the fact that the particle starts moving from asymptotic infinity ($r \rightarrow \infty$) which also corresponds to a large ``proper'' radial distance ($\rho \rightarrow \infty$). As a trivial exercise, this can be checked by setting $\eta=\eta_c$ a constant while taking $\sigma \rightarrow \infty$. Using the asymptotic function \eqref{e2.22}, this yields
\begin{align}
    \rho|_{\sigma \rightarrow \infty} =\sqrt{a_2}P^{1/16}V_0^{3/16}\int \frac{d\sigma}{(\sigma^2+\eta_c^2)^{7/32}}\simeq\frac{16 \sqrt{a_2}}{9}P^{1/16}V_0^{3/16}\sigma^{9/16}
\end{align}
which clearly diverges in the asymptotic ($\sigma \rightarrow \infty$) limit.

In other words, we place the massive probe close to the $N$ D0 branes sitting in the asymptotics and let it travel to the interior of the bulk along the proper radial direction $\rho$. The energy (or the Hamiltonian \eqref{e2.15}) of the particle is determined by placing the particle near the asymptotic infinity ($r\rightarrow \infty$), which yields
\begin{align}
\label{e2.24}
    H_0|_{r \rightarrow \infty}&=\sqrt{h_1(r)}\Big[ 1+\frac{\alpha}{r^7}(\dot{\sigma}^2+\dot{\eta}^2)\Big]^{1/2}~;~\alpha=\frac{a_2 P}{a_1 V_0}
\end{align}
where the sub-leading corrections are clearly suppressed considering the fact that the velocity of the particle in the asymptotic infinity is finite. In other words, the leading order term in the Hamiltonian \eqref{e2.24} dominates in the asymptotic limit ($r \rightarrow \infty$), which yields a large number. This would imply that we probe the BMN sector with a \emph{heavy} operator.

We are interested in exploring the size (or the rate of growth of complexity) associated with this heavy operator in the course of time. This is done by studying the trajectory of the massive particle in the ($\sigma ,\eta$) plane and along the proper radial direction ($\rho$). 

Using the Hamiltonian constraint \eqref{e2.24}, one finds the following combination
\begin{align}
\label{e2.25}
    \dot{\sigma}^2+\dot{\eta}^2\Big|_{r \rightarrow \infty}=-\frac{r^7}{\alpha}\Big(1-\frac{H^2_0}{h_1} \Big).
\end{align}

Notice that in the strict asymptotic limit ($r \rightarrow \infty$), $H^2_0 \sim h_1 \sim r^{49/8}$ \eqref{e2.24}, while the velocity can still be kept finite. This follows from the fact that the r.h.s. of \eqref{e2.25} is a product of a very small number with a large number ($\sim r^7$). This would correspond to a finite proper momentum \eqref{e2.20} and hence an increase in the size of the \emph{precursor}.

To calculate the proper momentum \eqref{e2.20}, we replace $|L|$ using \eqref{e2.15} and thereby take a large $r \rightarrow \infty$ limit. Using \eqref{e2.24}, this finally yields
\begin{align}
\label{e2.26}
    \mathcal{P}_\rho |_{r \rightarrow \infty} =\frac{\sqrt{\alpha}}{r^{7/2}}\sqrt{\dot{\sigma}^2+\dot{\eta}^2}+\mathcal{O}(r^{-21/2}).
\end{align}
\subsection{Early time growth of complexity}
In order to explore the early time behavior of the complexity ($\mathcal{C}(t)$), one has to solve the equations \eqref{e2.16}-\eqref{e2.17} near the asymptotics ($r\rightarrow \infty$) of the spacetime. 

Considering the fact that (where we calculate the combination $(\dot{\sigma}^2+\dot{\eta}^2)$ from \eqref{e2.15})
\begin{align}
    &\frac{\partial_\sigma h_1}{h_1}\Big |_{r \rightarrow \infty}=\frac{2 \sigma}{r^2}~;~\frac{\partial_\eta h_1}{h_1}\Big |_{r \rightarrow \infty}=\frac{2 \eta}{r^2}~;~\frac{h_2}{h_1}\Big|_{r \rightarrow \infty}=\frac{\alpha}{r^7}\\
    &\frac{\partial_\sigma h_2}{h_1}(\dot{\sigma^2}+\dot{\eta}^2)\Big|_{r \rightarrow \infty}=-\frac{7\sigma}{8r^2}+\frac{7a_1 V^{11/8}_0 }{8H^2_0}\frac{\sigma r^{33/8}}{P^{7/8}}\nonumber\\
    &\frac{\partial_\eta h_2}{h_1}(\dot{\sigma^2}+\dot{\eta}^2)\Big|_{r \rightarrow \infty}=-\frac{7 \eta}{8 r^2}+\frac{7a_1 V^{11/8}_0 }{8H^2_0}\frac{\eta r^{33/8}}{P^{7/8}}
\end{align}
the equations of motion \eqref{e2.16}-\eqref{e2.17} get simplified 
\begin{align}
\label{e2.29}
    &\frac{d}{dt}\Big[ \frac{\alpha \dot{\sigma}}{r^7}\Big]+\frac{23 \sigma}{16 r^2}-\frac{\beta \sigma }{H^2_0}r^{33/8} =0\\
    &\frac{d}{dt}\Big[ \frac{\alpha\dot{\eta}}{r^7}\Big]+\frac{23 \eta}{16 r^2}-\frac{\beta \eta }{H^2_0}r^{33/8} =0=0
    \label{e2.30}
\end{align}
where we have defined $\beta=\frac{7a_1 V^{11/8}_0}{16 P^{7/8}}$.

Using \eqref{e2.24}, one can show that at leading order in the large $r$ expansion
\begin{align}
\label{e2.31}
    \frac{\beta}{H^2_0}=\frac{7}{16}r^{-49/8}.
\end{align}

Multiplying \eqref{e2.29} by $\sigma$ and \eqref{e2.30} by $\eta$ and adding them together, we find the following
\begin{align}
\label{e2.32}
    \sigma \ddot{\sigma}+\eta \ddot{\eta}-7\dot{r}^2+\frac{23}{16}\frac{r^7}{\alpha}-\frac{\beta}{\alpha H^2_0}r^{105/8}=0.
\end{align}

Using \eqref{e2.31}, one can further simplify \eqref{e2.32} as
\begin{align}
\label{e2.33}
    \sigma \ddot{\sigma}+\eta \ddot{\eta}-7\dot{r}^2+\frac{r^7}{\alpha}=0.
\end{align}

Noting the fact that $r^2=\sigma^2+\eta^2$, we find the following combination
\begin{align}
\label{e2.34}
    \sigma \ddot{\sigma}+\eta \ddot{\eta}=r \ddot{r}+\dot{r}^2-(\dot{\sigma}^2+\dot{\eta}^2).
\end{align}

Multiplying \eqref{e2.29} by $\dot{\sigma}$ and \eqref{e2.30} by $\dot{\eta}$ and adding them together, one finds the following combination
\begin{align}
\label{e2.35}
    \frac{d}{dt}\Big[ (\dot{\sigma}^2+\dot{\eta}^2)+\frac{2r^7}{7 \alpha}\Big]=\frac{14 \dot{r}}{r}(\dot{\sigma}^2+\dot{\eta}^2).
\end{align}

Considering the fact that both $\dot{\sigma}\sim 0$ and $\dot{\eta}\sim 0$ are vanishingly small near the asymptotic infinity ($r \rightarrow \infty$), one finds the r.h.s. of \eqref{e2.35}
\begin{align}
  \frac{\dot{r}}{r}(\dot{\sigma}^2+\dot{\eta}^2) \Big|_{r \rightarrow \infty}=\frac{\sigma}{r^2}(\dot{\sigma }^3+\dot{\sigma}\dot{\eta}^2) +\frac{\eta}{r^2}(\dot{\eta}^3+\dot{\eta}\dot{\sigma}^2)\sim \frac{v^3}{r}\sim 0.
\end{align}

Integrating \eqref{e2.35}, we obtain
\begin{align}
\label{e2.37}
    \dot{\sigma}^2+\dot{\eta}^2=-\frac{2r^7}{7 \alpha}\Big(1-\frac{r^7_{UV}}{r^7} \Big)
\end{align}
where $r_{UV}$ is the UV cut-off.

Using \eqref{e2.37}, we finally simplify the $r(t)$ eq. \eqref{e2.33} to yield 
\begin{align}
\label{e2.38}
    r \ddot{r}-6 \dot{r}^2+\frac{9 r^7}{7 \alpha}-C_0=0
\end{align}
where the constant $C_0=\frac{2r_{UV}^7}{7 \alpha}$ is fixed in terms of the UV cut-off ($r_{UV}$).

In order to solve \eqref{e2.38}, we propose a solution in the form
\begin{align}
\label{e2.39}
    r(t)=r_{UV}(1- f(t))
\end{align}
such that the function $f(t)$ reflects a small deviation from the UV location at an early time $t \sim 0$. Substituting \eqref{e2.39} into \eqref{e2.38} and considering the terms linear in the fluctuation, we find the following linearized equation
\begin{align}
\label{e2.40}
    \ddot{f}-\gamma (1-9 f)\approx 0~;~\gamma = \frac{r^5_{UV}}{\alpha}.
\end{align}

The above Eq. \eqref{e2.40} has a solution
\begin{align}
    f(t)=\frac{1}{9} \Big(1-\cos \left(3 \sqrt{\gamma } t\right)\Big)
\end{align}
that vanishes at $t=0$ revealing $r(t=0)=r_{UV}$.

The complete solution can be expressed as
\begin{align}
\label{e2.42}
    r(t)=r_{UV}\Big[1-\frac{1}{9} \Big(1-\cos \left(3 \sqrt{\gamma } t\right)\Big)\Big].
\end{align}

The negative sign indicates that the particle starts to move towards lower values of $r$ with increasing time. Using \eqref{e2.37} and \eqref{e2.42}, we finally obtain the rate of growth of complexity (at early times) that follows from \eqref{e2.26}
\begin{align}
\label{e2.43}
    \frac{d \mathcal{C}}{dt}\Big|_{t \sim 0}=\mathcal{P}_\rho |_{t \sim 0} =\frac{\sqrt{a_1 V_0}}{\sqrt{a_2 P}}r^{5/2}_{UV}t \Rightarrow \mathcal{C}(t) =\frac{\sqrt{a_1 V_0}}{\sqrt{a_2 P}}r^{5/2}_{UV} t^2.
\end{align}
\subsection{Complexity with one conducting disk: D2 brane solution}
The above analysis gives an intuitive idea of growth of holographic complexity at very early time scales, and it lacks the full picture of time evolution of complexity in the matrix model. In order to have a complete picture, one has to solve the geodesic of the massive probe for an arbitrary value of the radial coordinate. 

The question we would like to ask is what will happen when the particle goes deeper inside the bulk or towards small values of $r$. Clearly, the particle will experience the presence of concentric spherical D2 or NS5 branes that are dual to fuzzy spheres (vacuum) in the matrix model. We address this question with a simplified gravity dual of PWMM that contains a finite conducting disk on top of the infinite conducting disk at $\eta =0$ \cite{Ling:2006up}-\cite{Asano:2014vba}. 

The corresponding potential function is given by
\begin{align}
\label{e2.44}
     V_{PWMM}(\sigma , \eta)=V_0 \Big(\eta \sigma^2-\frac{2}{3}\eta^3 \Big)+V_0 R^3 \phi_\kappa (\sigma/R , \eta/R).
\end{align}

The above potential corresponds to a finite conducting disk that is placed at $\eta=\eta_0>0$ with radius $R$ and carrying a charge $Q$. The corresponding NS5 and D2 brane charges are given by $N_{NS5}=\frac{2 \eta_0}{\pi}$ and $n_{D2}=\frac{8Q}{\pi^2}$. Combining them, we find the rank of the matrices $N=n_{D2} N_{NS5}=\frac{16 P}{\pi^3}$ where $P=Qd$ is the dipole moment of the disk. The parameter $\kappa$ is defined as the ratio $\kappa = \frac{\eta_0}{R}$. The function $\phi_\kappa(\sigma/R , \eta/R)$ can be expressed as \cite{Ling:2006up}-\cite{Asano:2014vba}
\begin{align}
    \phi_\kappa(\sigma/R , \eta/R)=\frac{\beta}{\pi}\int_{-1}^{1}dx \Bigg[-\frac{1}{\sqrt{\frac{\sigma^2}{R^2}+\Big(\frac{\eta}{R}+\kappa+i x \Big)^2}} +\frac{1}{\sqrt{\frac{\sigma^2}{R^2}+\Big(\frac{\eta}{R}-\kappa+i x \Big)^2}}\Bigg]f_\kappa (x).
\end{align}

Here, $f_\kappa (x)$ is the solution to the Fredholm integral equation of the second kind, and $\beta(\kappa)$ is given in terms of these functions. The function $f_\kappa$ is related to the charge density ($\varrho_\kappa (\sigma)$) on the disk as \cite{Ling:2006up}
\begin{align}
    &f_\kappa (x)=\frac{2\pi}{\beta}\int_x^1 d\sigma\frac{\sigma \varrho_\kappa (\sigma)}{(\sigma^2 -x^2)^{1/2}}\\
    &\varrho_\kappa (\sigma)=\frac{\beta}{\pi^2}\Big[ \frac{f_\kappa (1)}{(1-\sigma^2)^{1/2}}-\int_\sigma^1 dx\frac{f_\kappa'(x)}{(x^2-\sigma^2)^{1/2}}\Big].
\end{align}
In the following, we discuss various limits of the supergravity solution that results from the potential \eqref{e2.44} and calculate the complexity in each of those cases. 

The D2 brane limit corresponds to a finite conducting size disk at $\eta=0$ carrying a charge $Q=\frac{\pi^2 n_{D2}}{8}$, while keeping $Q$ (or equivalently, the radius $R$) and the ratio $\frac{g^2}{N_{NS5}}=g^{2}_{S^2}$ fixed in the limit $N_{NS5} \rightarrow \infty$. This sets the parameter $\kappa \rightarrow \infty$. The PWMM eventually boils down to $\mathcal{N}= 8$ SYM in $R \times S^2$, where $g^{2}_{S^2}$ is identified as the gauge coupling of the $\mathcal{N}= 8$ SYM theory. In the following, we outline the solution in the line of the papers \cite{Lin:2005nh}, \cite{Ling:2006up}. The general form of the potential for single disk D2 brane solution is given by
\begin{align}
\label{e2.48}
    V(\sigma, \eta)=W_0(\sigma^2-2\eta^2)+\phi(\sigma , \eta)=V_{D2}(\sigma , \eta)+\phi (\sigma , \eta)
\end{align}
where $V_{D2}$ is the background potential for a charge conducting disk of charge $Q$ and radius $R$. We set $W_0=V_0 \eta_0$ finite ($=1$) in the limit $\eta_0 \rightarrow \infty$ and also set $R=1$.

The potential \eqref{e2.48} must approach $V_{D2}$ in the asymptotic limit, which corresponds to the near horizon geometry of D2 branes wrapping a two sphere. In other words, $\phi$ must vanish at infinity. On the other hand, on the disk one finds
\begin{align}
    \phi (\sigma , \eta=0)=2-\sigma^2
\end{align}
which ensures a vanishing charge density at the edges of the disk.

The complete 10d solution of the single disk D2 brane configuration has been obtained by the authors in \cite{Lin:2005nh}. For our purpose, we note down the metric and the dilaton 
\begin{align}
    &ds^2_{string}=-g_1(r,\theta)dt^2+g_2(r,\theta)dr^2+g_3(r, \theta)d\theta^2+g_4(r, \theta)d\Omega^2_2+g_5(r, \theta)d\Omega^2_5\\
    &g_1(r,\theta)=8(1+r^2)f(r,\theta)~;~f(r,\theta)=\sqrt{\frac{2}{r}}\sqrt{r+(\cos^2\theta+r^2)\arctan r}\\
    &g_2(r, \theta)=\frac{8 r f(r, \theta)}{r+(1+r^2)\arctan r}\frac{1}{(1+r^2)}~;~g_3(r, \theta)=(1+r^2)g_2(r, \theta)\\
    &g_4(r,\theta)=\frac{2r f(r,\theta)}{1+r \arctan r}(r+(1+r^2)\arctan r)~;~g_5(r,\theta)=16 f^{-1}(r, \theta)\sin^2\theta\\
    &e^{\phi}=\frac{8 \sqrt{r}}{\sqrt{1+r \arctan r}}\frac{1}{\sqrt{f(r , \theta)}}\frac{1}{\sqrt{r+(1+r^2)\arctan r}}
\end{align}
where the solution has been expressed in a different set of coordinates $(\sigma ,\eta)\rightarrow (r,\theta)$.

We parametrize the geodesic of the particle by choosing $r=r(t)$ and $\theta = \theta(t)$ while all remaining coordinates (on two sphere and five sphere) are held fixed. This leads to the geodesic of the massive probe in Einstein's frame as
\begin{align}
    &S_p=\int dt L~;~L=-\sqrt{h_1(r,\theta)-h_2(r,\theta)\dot{r}^2-h_3(r,\theta)\dot{\theta}^2}\\
    &h_1(r,\theta)=\frac{\sqrt{8}(1+r^2)f^{5/4}(r, \theta)}{r^{1/4}}(1+r \arctan r)^{1/4}(r+(1+r^2)\arctan r)^{1/4}\\
    &h_2(r, \theta)=\frac{\sqrt{8}r^{3/4}f^{5/4}(r,\theta)}{(1+r^2)}\frac{(1+r \arctan r)^{1/4}}{(r+(1+r^2)\arctan r)^{3/4}}\\
    &h_3(r, \theta)=\sqrt{8}r^{3/4}f^{5/4}(r,\theta)\frac{(1+r \arctan r)^{1/4}}{(r+(1+r^2)\arctan r)^{3/4}}.
\end{align}

The equations of motion can be expressed as
\begin{align}
\label{e2.59}
   &- \frac{d}{dt}\Big[\frac{h_2}{L}\dot{r} \Big]=\frac{1}{2L}(\partial_r h_1 -\partial_r h_2 \dot{r}^2 -\partial_rh_3 \dot{\theta}^2)\\
   &- \frac{d}{dt}\Big[\frac{h_3}{L}\dot{\theta} \Big]=\frac{1}{2L}(\partial_\theta h_1 -\partial_\theta h_2 \dot{r}^2 -\partial_\theta h_3 \dot{\theta}^2).
   \label{e2.60}
\end{align}

Notice that $\partial_\theta h_i \sim \partial_\theta f \sim \sin \theta \cos\theta$. Therefore, $\ddot{\theta}=\dot{\theta}=\theta=0$ is a solution of the $\theta$- equation of motion \eqref{e2.60}. Imposing this on the $r$- equation of motion \eqref{e2.59}, we obtain
\begin{align}
\label{e2.61}
    - \frac{d}{dt}\Big[\frac{\hat{h}_2}{\hat{L}}\dot{r} \Big]=\frac{1}{2\hat{L}}(\partial_r \hat{h}_1 -\partial_r \hat{h}_2 \dot{r}^2 )
\end{align}
where we denote the above functions as
\begin{align}
    &\hat{L}=-\sqrt{\hat{h}_1(r)-\hat{h}_2(r)\dot{r}^2}\\
    &\hat{h}_1(r)=\frac{2^{17/8}}{r^{7/8}}(1+r^2)(r+(1+r^2)\arctan r)^{7/8}(1+r \arctan r)^{1/4}\\
    &\hat{h}_2(r)=\frac{2^{17/8}r^{1/8}}{(1+r^2)}\frac{(1+r \arctan r)^{1/4}}{(r+(1+r^2)\arctan r)^{1/8}}.
\end{align}

The corresponding Hamiltonian is given by
\begin{align}
\label{e2.65}
    H_0 = \frac{\hat{h}_1(r)}{|\hat{L}|}.
\end{align}

Like before, we can fix the Hamiltonian by setting $\dot{r}|_{r=r_{UV}}=0$, which yields $H_0 =\sqrt{2} \pi ^{9/16} r_{UV}^{25/16} $, which is a large number that corresponds to a heavy operator inserted in $\mathcal{N}=8$ SYM theory. Using \eqref{e2.65}, we can simplify the equation of motion \eqref{e2.61} to obtain
\begin{align}
\label{e2.66}
    - \frac{d}{dt}\Big[\frac{\hat{h}_2}{\hat{h}_1}\dot{r} \Big]=\frac{1}{2\hat{h}_1}(\partial_r \hat{h}_1 -\partial_r \hat{h}_2 \dot{r}^2 ).
\end{align}

\begin{figure}
    \centering
    \includegraphics[width=0.9\linewidth]{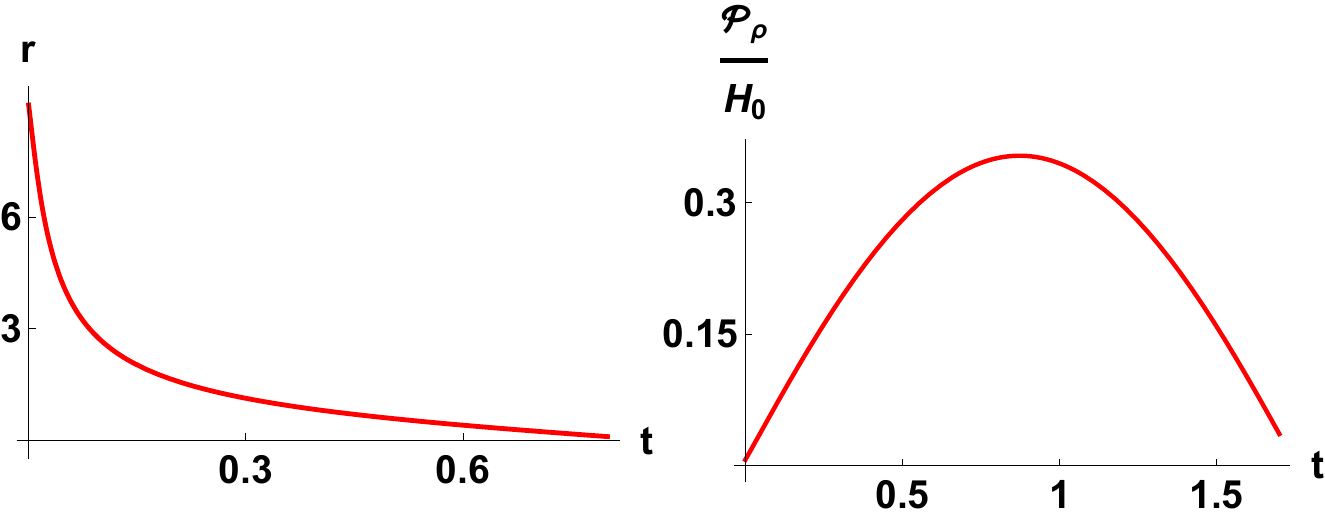}
    \caption{We plot the trajectory $r(t)$ and the proper momentum $\mathcal{P}_\rho (t)$ of the massive probe as a function of time (t). These plots are obtained by integrating \eqref{e2.67} and evaluating \eqref{e2.69} on the solution. We choose the initial condition as $r_{UV}=10$ and $r(t=0)=r_0=9$.}
    \label{figd2}
\end{figure}

\begin{figure}
    \centering
    \includegraphics[width=0.5\linewidth]{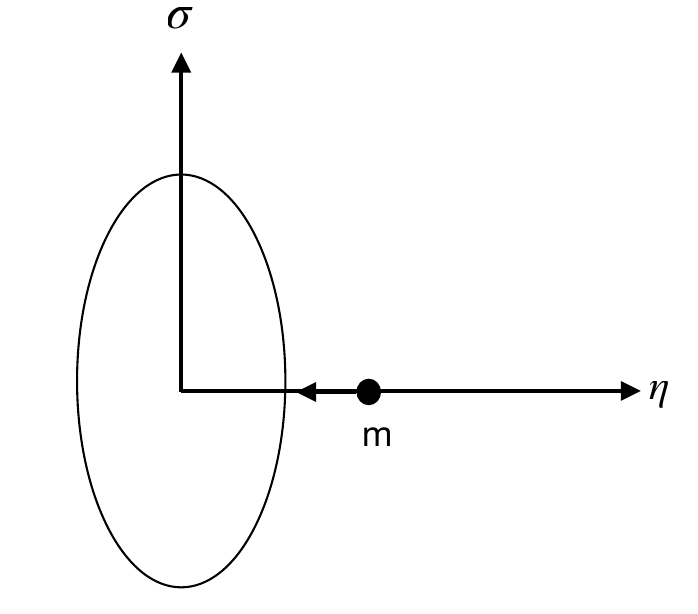}
    \caption{We plot the particle trajectory in the D2 brane limit. The trajectory (or the configuration) corresponds to a physical solution to the equations of motion for $\theta =0$.}
    \label{figd2p}
\end{figure}

\begin{figure}
    \centering
    \includegraphics[width=0.5\linewidth]{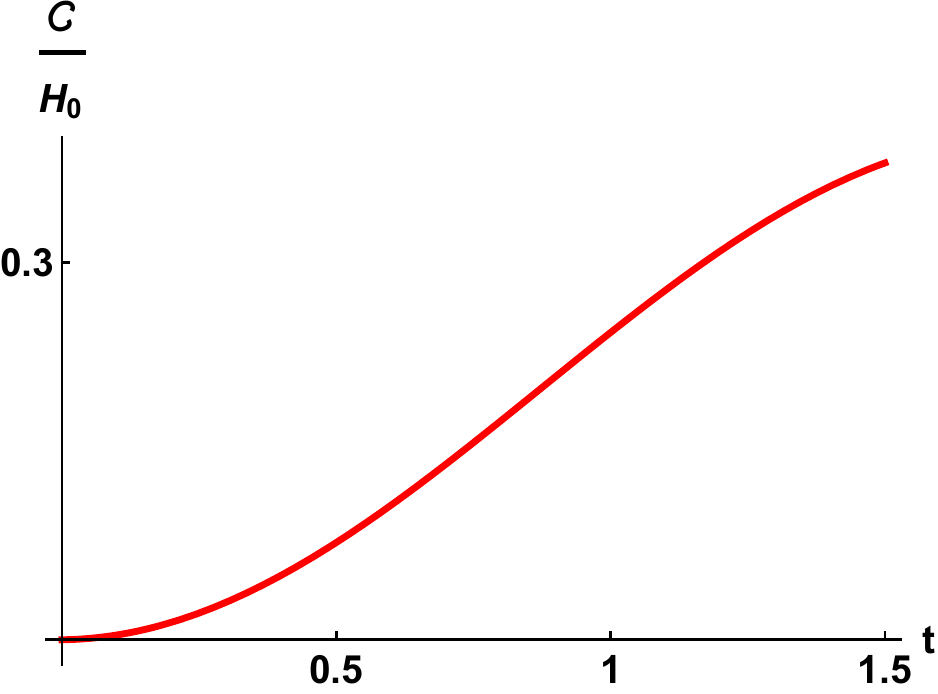}
    \caption{We plot complexity $\mathcal{C}(t)$ as a function of time (t). We choose the initial condition as $r_{UV}=10$ and $r(t=0)=r_0=9$.}
    \label{figcd2}
\end{figure}

It is straightforward to show that the equation of motion \eqref{e2.66} is equivalent to $\frac{dH_0}{dt}=0$. In other words, we solve the constraint \eqref{e2.65} to obtain the trajectory $r(t)$ of the particle
\begin{align}
\label{e2.67}
    \dot{r}(t)=-\sqrt{\frac{\hat{h}_1}{\hat{h}_2}}\sqrt{1-\frac{\hat{h}_1}{H^2_0}}.
\end{align}

Notice that in the strict UV limit $\hat{h}_1 (r)|_{r=r_{UV}}=H^2_0$, which yields a zero initial velocity and is therefore consistent with our assumption. On the other hand, for $r< r_{UV}$, one can expand the r.h.s. under the square root in $1/H_0^2$ to yield a LO solution. The negative sign indicates that $r(t)$ decreases with increasing time (Fig.\ref{figd2}).

The proper momentum ($\mathcal{P}_\rho$) can be defined in terms of the proper distance \cite{Caputa:2024sux}-\cite{Fatemiabhari:2025cyy}
\begin{align}
\label{e2.68}
    d\rho = \sqrt{\hat{h}_2 (r)}dr.
\end{align}

Using \eqref{e2.68}, the proper momentum can be expressed as
\begin{align}
\label{e2.69}
  \frac{d \mathcal{C}}{dt}=  \mathcal{P}_\rho = \frac{\partial \hat{L}}{\partial \dot{\rho}}=H_0 \frac{\sqrt{\hat{h}_2}}{\hat{h}_1}\dot{r}= \frac{H_0}{\sqrt{\hat{h}_1}}\sqrt{1-\frac{\hat{h}_1}{H^2_0}}
\end{align}
where $H_0$ plays the role of the UV cut-off \cite{Caputa:2024sux}.

It turns out that in terms of the original coordinates ($\sigma , \eta$), the choice $\theta=0$ corresponds to particle motion along the $\sigma =0$ axis \cite{Lin:2005nh}. At late times the particle approaches $r \sim 0$, which is equivalent to $\eta \sim 0$. This is precisely the location of the conducting disk in the bulk (Fig.\ref{figd2p}). In terms of the proper distance \eqref{e2.68}, one can show that $\rho \sim 2 r \sim 2 \eta \sim 0$. In other words, the particle falls from infinity ($r \sim r_{UV}$) and is reflected back by the conducting disk placed at $\eta =0$. This causes a deflection in the proper momentum as shown in Fig.\ref{figd2}.

From the particle's frame, it interpolates between a non-AdS asymptotic 
\begin{align}
    ds^2_{UV}\sim -2 \pi ^{9/8} r^{25/8} dt^2+4 \sqrt[8]{\pi } \left(\frac{1}{r}\right)^{15/8}dr^2
\end{align}
and a two dimensional Minkowski in the IR (after a suitable rescaling of time)
\begin{align}
    ds^2_{IR}\sim 4(- dt^2 + dr^2).
\end{align}

Finally, the complexity of the dual operator can be obtained by integrating \eqref{e2.69}
\begin{align}
    \mathcal{C}/H_0 = \int dt \mathcal{P}_\rho (t).
\end{align}

As Fig.\ref{figcd2} reveals, the complexity initially grows quadratically with time and thus reaches a maximum when the particle approaches the disk near $\rho \sim r \sim 0$. Quadratic growth is an artifact of the linear growth of the proper momentum ($\mathcal{P}_\rho$) at the initial time. Later, the complexity starts to saturate with decreasing momentum as the particle approaches UV.
\subsection{Two infinite conducting plates: NS5 brane solution}
The NS5 brane limit corresponds to setting $Q \rightarrow \infty$ while keeping $N_{NS5}$ fixed. In the electrostatic description, this corresponds to two infinite conducting planes separated by a fixed distance $N$, where $N$ corresponds to the number of NS5 branes wrapping $R \times S^5$. 

The metric and the dilaton relevant for our analysis are given by \cite{Lin:2005nh}
\begin{align}
\label{e2.73}
    & ds^2_{string}=-k_1(r)dt^2+k_2(r)(dr^2+d \theta^2)+k_3(r, \theta)d\Omega^2_2+k_4(r)d\Omega^2_5\\
    & k_1 (r)=2  N r\sqrt{\frac{I_0}{I_2}}~;~k_2(r)=N \frac{\sqrt{I_2 I_0}}{I_1}\\
    & k_3(r, \theta)=\frac{\sqrt{I_0 I_2}I_1 \sin^2\theta}{I_0 I_2 \sin^2\theta + I^2_1 \cos^2\theta}~;~k_4(r)=2N r \sqrt{\frac{I_2}{I_0}}\\
    &e^{\phi}=\frac{g_0 N^{3/2}}{2}\Big( \frac{I_2}{I_0}\Big)^{3/4}\sqrt{\frac{I_0}{I_1}}(I_0 I_2 \sin^2 \theta +I^2_1 \cos^2 \theta)^{-1/2}
\end{align}
where $I_n (r)$ is the modified Bessel functions of the first kind.

The solution \eqref{e2.73} in the UV ($r \rightarrow \infty$) corresponds to the near horizon geometry of $N$ NS5 branes wrapping a five sphere ($S^5$). The corresponding dual theory is known as the Little String Theory on $S^5$, which is a non-gravitational theory in six dimensions \cite{Ling:2006up}.

Like before, we choose a geodesic in the Einstein's frame which is parametrized by $r=r(t)$ and $\theta = \theta (t)$, while all the remaining coordinates of the internal space are held fixed. This results in the following action for the massive probe
\begin{align}
    &S_p=\int dt L ~;~ L=-\sqrt{n_1(r, \theta)-n_2(r,\theta)(\dot{r}^2+\dot{\theta}^2)}\\
    & n_1(r, \theta)=\frac{2^{3/2}}{\sqrt{g_0}}N^{1/4}r \Big( \frac{I_0}{I_2}\Big)^{7/8}\Big( \frac{I_1}{I_0}\Big)^{1/4}G(r, \theta)^{1/4}~;~G(r, \theta)=I_0 I_2 \sin^2\theta + I^2_1 \cos^2\theta\\
    &n_2 (r, \theta)=\sqrt{\frac{2}{g_0}}N^{1/4}\Big( \frac{I_2}{I_0}\Big)^{1/8}\Big( \frac{I_0}{I_1}\Big)^{3/4}G(r, \theta)^{1/4}.
\end{align}
\subsubsection{Configuration with $\theta =0$}
The equation of motion for $\theta (t)$ has a trivial solution $\ddot{\theta}=\dot{\theta}=\theta =0$. In the original electrostatic coordinates ($ \sigma , \eta$), this would correspond to the motion of the test probe along the $\eta =0$ axis, while $\sigma \in [- \infty , \infty]$ remains unbounded.  

This further simplifies the Lagrangian of the point particle action to yield
\begin{align}
    &S_p |_{\theta =0}=\int dt \hat{L} ~;~ \hat{L}=-\sqrt{\hat{n}_1(r)-\hat{n}_2(r)\dot{r}^2}\\
    & \hat{n}_1(r)=\frac{2^{3/2}}{\sqrt{g_0}}N^{1/4}r \sqrt{I_1}\Big( \frac{I_0}{I_2}\Big)^{7/8}\Big( \frac{I_1}{I_0}\Big)^{1/4}\\
    &\hat{n}_2 (r)=\sqrt{\frac{2}{g_0}}N^{1/4}\sqrt{I_1}\Big( \frac{I_2}{I_0}\Big)^{1/8}\Big( \frac{I_0}{I_1}\Big)^{3/4}.
\end{align}

The corresponding equation of motion can be expressed as
\begin{align}
\label{e2.83}
    - \frac{d}{dt}\Big[\frac{\hat{n}_2}{\hat{L}}\dot{r} \Big]=\frac{1}{2\hat{L}}(\partial_r \hat{n}_1 -\partial_r \hat{n}_2 \dot{r}^2 ).
\end{align}

\begin{figure}
    \centering
    \includegraphics[width=0.9\linewidth]{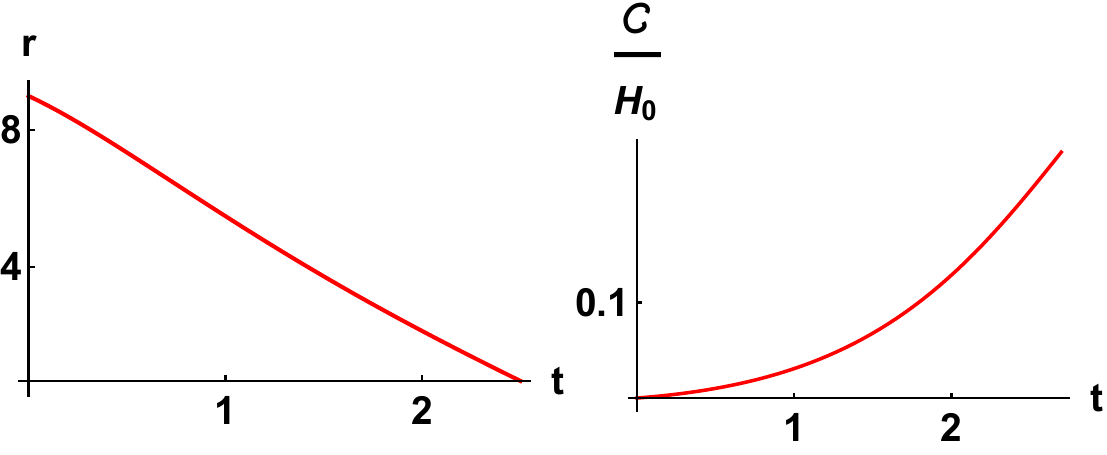}
    \caption{We plot the trajectory $r(t)$ of the particle and the complexity $\mathcal{C} (t)$ of the dual operator as a function of time (t). These plots are obtained by integrating \eqref{e2.84} and evaluating \eqref{e2.85} on the solution. We choose the initial condition as $N=100$, $g_0=2$, $r_{UV}=10$ and $r(t=0)=r_0=9$.}
    \label{fign5}
\end{figure}

\begin{figure}
    \centering
    \includegraphics[width=0.5\linewidth]{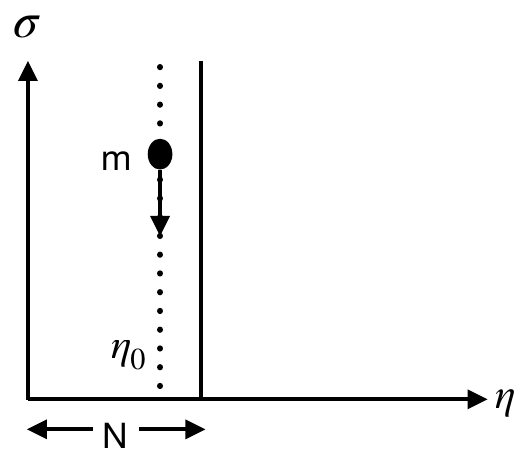}
    \caption{Particle trajectory corresponding to $\theta= \frac{\pi}{2}$ in the NS5 brane limit. This is a physical configuration of the massive probe, allowed by the dynamics of the system.}
    \label{fign5p}
\end{figure}

The rest of the discussion goes identical to the D2 brane limit as in the previous section. We therefore summarize the key results below, namely the trajectory ($r(t)$) of the particle and the proper momentum ($\mathcal{P}_\rho$)
\begin{align}
\label{e2.84}
    &\dot{r}(t)=-\sqrt{\frac{\hat{n}_1}{\hat{n}_2}}\sqrt{1-\frac{\hat{n}_1}{H^2_0}}~;~H_0 = \frac{\hat{n}_1(r)}{|\hat{L}|}\\
    & \dot{\mathcal{C}}=\mathcal{P}_\rho = \frac{H_0}{\sqrt{\hat{n}_1}}\sqrt{1-\frac{\hat{n}_1}{H^2_0}}.
    \label{e2.85}
\end{align}

Before we compute the proper momentum and complexity, it is customary to explore the behaviors of the functions $\hat{n}_1(r)$ and $\hat{n}_2(r)$ in the large and small $r$ limits
\begin{align}
   \hat{n}_1(r)|_{r \rightarrow \infty}=\frac{2 N^{1/4}r^{3/4}}{(2 \pi)^{1/4}}e^{r/2} ~;~ \hat{n}_2(r)|_{r \rightarrow \infty}=\frac{N^{1/4}}{(2 \pi)^{1/4}}\frac{e^{r/2}}{r^{1/4}}
\end{align}
where we set $g_0=2$ for simplicity. 

The UV geometry that the particle probes appear to be (following a rescaling $t \rightarrow \sqrt{2}t$)
\begin{align}
\label{e2.87}
    ds^2|_{UV}=\Omega(r)\Big[-dt^2 +\frac{dr^2}{r} \Big]~;~\Omega(r)=\frac{N^{1/4}}{(2 \pi)^{1/4}}r^{3/4}e^{r/2}
\end{align}
which is conformally equivalent to a non-AdS geometry. The corresponding proper momentum scales as $\mathcal{P}_\rho/H_0 \sim r^{-3/8}e^{-r/4}\sim 0$, which yields a zero rate of growth of complexity.

On a similar note, the functions near $r \sim 0$ behave as
\begin{align}
\label{e2.88}
    \hat{n}_1(r)|_{r \sim 0}\sim 1~;~\hat{n}_2(r)|_{r \sim 0}\sim 1 \Rightarrow ds^2|_{IR}\sim -dt^2 + dr^2.
\end{align}
In other words, the particle experiences a 2d Minkowski geometry in the deep IR. 

    Clearly, as Fig.\ref{fign5} reveals, as time progresses, the particle moves towards the interior ($r \sim 0$) of the bulk starting from some position ($r_0$) near the UV ($r_{UV}$). In the original electrostatic coordinates ($\sigma ,\eta$), this refers to a motion towards smaller values of $\sigma \sim 0$ along the $\eta =0$ axis. The corresponding complexity ($\mathcal{C}(t)$) plot reveals a non-linear growth to begin with, which further increases at later times (see Fig.\ref{fign5}). Compared with the D2 brane solution (Fig.\ref{figcd2}), one notices a basic qualitative difference, namely the complexity in the D2 brane limit saturates faster than the NS5 brane configuration.

\subsubsection{Configuration with $\theta =\frac{\pi}{2}$ } 
Notice that in the above configuration, the particle does not feel any potential ($V=0$) as it travels along the $\theta =0$ axis, which refers to the infinite conducting disk at $\eta=0$. Therefore, to see the effects of the other conducting disk, one has to find a configuration with non-zero $\theta$, which would correspond to a motion of the particle in the $(\sigma ,\eta)$ plane. 

Looking back at the $\theta$- equation of motion, we find
\begin{align}
\label{e2.89}
    -\frac{d}{dt}\Big[\frac{n_2 \dot{\theta}}{L} \Big]=\frac{1}{2 L}(\partial_\theta n_1 - \partial_\theta n_2 (\dot{r}^2 + \dot{\theta}^2)).
\end{align}

Notice that $\partial_\theta n_i \sim  \sin \theta \cos \theta/G(r , \theta)^{3/4}$, which clearly vanishes for $\theta =\pi/2$. In other words, this is the constant non-zero value of $\theta$ that solves the $\theta$ equation of motion \eqref{e2.89}. This would correspond to the motion of the massive probe between the two infinite conducting plates and at a fixed distance $\eta=\eta_0 = \frac{\pi N}{4}<N$ along the $\eta$-axis (Fig.\ref{fign5p}).

Given this configuration, the particle feels a potential $V \sim I_0(r)$ between the plates. The corresponding action and the Lagrangian density reads as (we will set $g_0=2$ as before)
\begin{align}
    &S_p |_{\theta =\frac{\pi}{2}}=\int dt \hat{L} ~;~ \hat{L}=-\sqrt{\hat{n}_1(r)-\hat{n}_2(r)\dot{r}^2}\\
    & \hat{n}_1(r)=\frac{2^{3/2}}{\sqrt{g_0}}N^{1/4}r \frac{I_0^{7/8}I_1^{1/4}}{I_2^{5/8}}\\
    &\hat{n}_2 (r)=\sqrt{\frac{2}{g_0}}N^{1/4}\frac{I^{3/8}_2 I_0^{7/8}}{I^{3/4}_1}.
\end{align}

\begin{figure}
    \centering
    \includegraphics[width=0.8\linewidth]{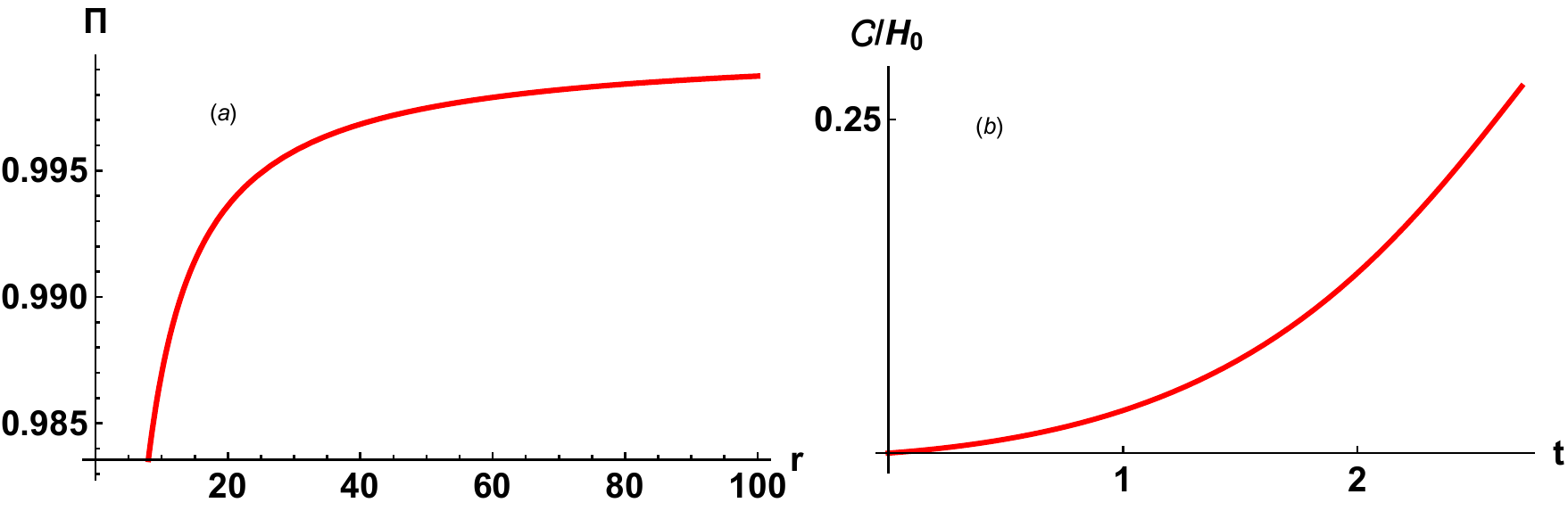}
    \caption{(a) We plot the ratio \eqref{e2.94}, that compares the rate of complexity growths for two different $\theta$- configurations. (b) We plot complexity for $\theta =\frac{\pi}{2}$ configuration.}
    \label{figpiC}
\end{figure}

In order to explore the geometry the particle experiences as it moves from the UV to $r\sim 0$, one has to expand the functions $\hat{n}_1(r)$ and $\hat{n}_2(r)$ both in the large $r$ and small $r$ limits. The geometry appears to be identical as in the previous example (\eqref{e2.87} and \eqref{e2.88}). This can be further confirmed by a closure comparison of the functions $\hat{n}_1(r)$ and $\hat{n}_2(r)$ for $\theta =0$ and $\theta =\frac{\pi}{2}$. For example, taking the ratio, one finds
\begin{align}
    \frac{\hat{n}_1|_{\theta=0}}{\hat{n}_1|_{\theta=\frac{\pi}{2}}}=\frac{I_1^{1/2}}{I_2^{1/4}I_0^{1/4}}~;~\frac{\hat{n}_2|_{\theta=0}}{\hat{n}_2|_{\theta=\frac{\pi}{2}}}=\frac{I_1^{1/2}}{I_2^{1/4}I_0^{1/4}}.
\end{align}
Clearly, these ratios become identical both in the limit $r \rightarrow \infty$ (where the function $I_n(r)\sim e^r/\sqrt{r}$) and in $r \sim 0$, where $I_n(r)\sim 1$. Since the functions $\hat{n}_i(r)$ appear to be (qualitatively) identical, hence the qualitative features of the complexity do not change for $\theta =\frac{\pi}{2}$.

In principle, one has to solve identical sets of equations \eqref{e2.84}-\eqref{e2.85} to find the trajectory ($r(t)$) of the particle and the complexity ($\mathcal{C}(t)\sim \int dt \mathcal{P}_\rho$). As usual, this is subjected to the Hamiltonian constraint which is obtained by setting the initial velocity equal to zero $\dot{r}(t=0)=0$, $H_0=\sqrt{\hat{n}_1(r)}|_{r=r_{UV}}\sim r_{UV}^{3/8}e^{r_{UV}/4}$, which clearly diverges near the boundary ($r_{UV}\rightarrow \infty$). A clousre comparison between the rate of growth of complexity for both ($\theta =0$ and $\theta =\frac{\pi}{2}$) configurations reveals the following ratio
\begin{align}
\label{e2.94}
    \Pi= \frac{\dot{\mathcal{C}}|_{\theta =0}}{\dot{\mathcal{C}}|_{\theta =\frac{\pi}{2}}}=\frac{I^{1/8}_2 I^{1/8}_0}{I^{1/4}_1}
\end{align}
where $1/H^2_0$ corrections are identical in UV and can be ignored for $r \sim 0$.

As Fig.\ref{figpiC}(a) reveals, the rate of growth of complexities in the UV ($r \rightarrow \infty$) are the same ($\Pi \sim 1$) for both $\theta$ configurations. On the other hand, they differ in the IR, $\Pi <1$. That is, the rate of growth of complexity for the $\theta =0$ configuration is less than that of the $\theta = \frac{\pi}{2}$ configuration. This is further evident from the complexity plot Fig.\ref{figpiC}(b).

\section{Lin solution of BMN matrix model and Krylov complexity}
\label{section 3}
Here we work out an example of the Lin solution \cite{Lin:2004kw} of constructing D0 branes that are dual to the deformations of the BFSS matrix model \cite{Banks:1996vh}. These are the type IIA solutions characterizing the near horizon geometry of N D0 branes perturbed by the background RR six form and NS-NS three form fluxes. This is an equivalent approach based on Polchinski and Strassler \cite{Polchinski:2000uf} to study the BMN matrix model, which is described as an electrostatic problem in the bulk supergravity description in the previous section.
\subsection{Complexity in the asymptotic limit}
Following \cite{Lin:2004kw}, we discuss complexity in the super-gravity limit of D2 brane configurations which is valid in the weak effective coupling limit of the matrix perturbation theory of the BMN matrix model \cite{Berenstein:2002jq}. The asymptotic solution ($r \rightarrow \infty$) corresponds to the metric of (near horizon limit of) N D0 branes perturbed by background RR and NS fluxes (those are dual to massive deformations of BFSS matrix model \cite{Banks:1996vh}) and are independent of the specific brane configurations in the small $r \sim 0$ regime. 

The metric and the dilaton in the large $r$ limit can be expressed as
\begin{align}
\label{e3.1}
    &ds_{string}^2=-h^{1/2}dt^2+h^{1/2}dr^2+h^{1/2}r^2 d\Omega^2_8\\
    &e^{\phi}=h^{3/4}~;~h(r)=\frac{R^7}{r^7}.
\end{align}

The corresponding metric in the Einstein frame can be expressed as
\begin{align}
    ds^2_E=-h_1(r)dt^2+h_2(r)dr^2+h_3(r)d\Omega^2_8
\end{align}
where the functions for the smeared Lin solution are given by
\begin{align}
    h_1(r)=\frac{r^{49/8}}{R^{49/8}}~;~h_2(r)=\frac{R^{7/8}}{r^{7/8}}~;~h_3(r)=R^{7/8}r^{9/8}.
\end{align}
\subsubsection{Point particle dynamics}
Notice that the large $r$ scalings of $h_1(r)$ and $h_2(r)$ are precisely those given by \eqref{e2.21} and \eqref{e2.22} respectively. We parametrize the geodesic of the massive probe with a choice $r=r(t)$ while keeping the coordinates of the eight sphere constant. This results in the following action of the massive particle (we set $m=1$ as before)
\begin{align}
    S_p=\int dt L ~;~L=-\sqrt{h_1(r)-h_2(r)\dot{r}^2}.
\end{align}

The canonical momentum is given by
\begin{align}
\label{e3.6}
    P_r = \frac{h_2(r)\dot{r}}{|L|}.
\end{align}

The Hamiltonian can be expressed as
\begin{align}
\label{e3.7}
    H_0=\frac{h_1(r)}{|L|}.
\end{align}

Like before, the Hamiltonian can be fixed by noting the fact that $\dot{r}|_{r=r_{UV}}=0$, where $r_{UV}\gg 1$ is the UV cut-off of our calculation. This yields
\begin{align}
    H_0 = \sqrt{h_1(r)}|_{r=r_{UV}}=\Big(\frac{r_{UV}}{R}\Big)^{49/16}.
\end{align}

The equation of motion for $r(t)$ reads as
\begin{align}
\label{e3.9}
    -\frac{d}{dt}\Big[ \frac{h_2(r)}{|L|}\dot{r}\Big]=\frac{1}{2|L|}(\partial_r h_1-\partial_r h_2 \dot{r}^2).
\end{align}

Using \eqref{e3.7}, we re-express \eqref{e3.9} as
\begin{align}
\label{e3.10}
    r \ddot{r}-\frac{105}{16}\dot{r}^2+\frac{49}{16}\frac{r^7}{R^7}=0
\end{align}
which is qualitatively similar to \eqref{e2.38}.

In order to solve \eqref{e3.10}, we propose an expansion near the UV scale
\begin{align}
\label{e3.11}
    r(t)=r_{UV}(1-f(t))
\end{align}
such that $|f(t)|\ll 1$ in the limit $t \sim 0$ and vanishes exactly at $t=0$. 

Keeping terms up to linear order in the fluctuation, we obtain
\begin{align}
    \ddot{f}(t)-\lambda \Big(1-7 f(t)\Big)\approx 0~;~\lambda = \frac{49}{16}\frac{r^5_{UV}}{R^7}.
\end{align}

The general solution is of the form
\begin{align}
    f(t)=\frac{1}{7}+c_2 \sin \left(\sqrt{7 \lambda}  t\right)+c_1 \cos \left(\sqrt{7 \lambda}  t\right).
\end{align}

The constants $c_1$ and $c_2$ are fixed from the boundary conditions, namely (i) $r(t=0)=r_{UV}$ and (ii) $\dot{r}|_{t=0}=0$. This fixes the constants as $c_2=0$ and $c_1=-\frac{1}{7}$. This leads to the complete solution \eqref{e3.11}, which is of the form
\begin{align}
    r(t)=r_{UV}\Big[ 1-\frac{1}{7}\Big( 1-\cos \left(\sqrt{7 \lambda}  t\right)\Big)\Big].
\end{align}
\subsubsection{Proper momentum and complexity}
The size of the operator in the BMN matrix model is conjectured to be dual to the proper momentum of the massive particle along the geodesic \cite{Caputa:2024sux}. Proper momentum comes with proper distance, which for the present model yields
\begin{align}
    \rho =\int dr \sqrt{h_2(r)}=\frac{16}{9}R^{7/16}r^{9/16}.
\end{align}

The corresponding proper momentum is given by
\begin{align}
    \mathcal{P}_\rho =-P_r \frac{\partial \dot{r}}{\partial \dot{\rho}}=-\Big(\frac{r_{UV}}{R}\Big)^{49/16}\frac{\sqrt{h_2}}{h_1}\dot{r}.
\end{align}

Notice that, as $\dot{r}(t)<0$, therefore, the negative sign is put by hand to make the size of the operator positive. After some calculations, one finally obtains
\begin{align}
\label{e3.17}
\frac{d \mathcal{C}}{dt}\Big|_{t \sim 0}=\mathcal{P}_\rho |_{t \sim 0} =\frac{49}{16}\Big( \frac{r_{UV}}{R}\Big)^{5/2}\frac{t}{R}
\end{align}
which reveals a linear growth in complexity at an early time.
\subsection{Complexity near the shell of D2 brane}
Clearly, the above calculation is performed in the domain of large $r$ and therefore is valid only at small time scales. We now extend the above calculation in the domain where the massive probe experiences the presence of spherical shells carrying D2 brane charges. Near the shell of the D2 branes, the presence of D0 charge ($N$) can be ignored, and the solution can be approximated by the near horizon limit of the $p$ flat D2 branes.

Consider a shell of concentric $p$ D2 branes carrying a charge $q$ each, thus giving a total D2 charge $n_{D2}=p q$. The total D0 charge is the sum of charges in the individual D2 shell, namely $N=\sum_i p_i q_i$, where $i$ stands for the $i$th shell. 

The metric and the dilaton near the shell of D2 brane is given by \cite{Lin:2004kw}
\begin{align}
\label{e3.18}
    &ds^2=-\frac{\sqrt{10}r_0 r^{5/2}}{R^{7/2}}dt^2+\frac{R^{7/2}}{\sqrt{10}r_0 r^{5/2}}(dr^2+r^2 d\Omega^2_6)+\frac{r^{5/2}R^{7/2}}{\sqrt{10}r_0}\frac{(dx^2_1+dx^2_2)}{(r^5+r_c^5)}\\
    &e^{-\Phi/2}=\frac{10^{3/8}r^{3/4}_0 r^{5/8}}{\sqrt{g_s}R^{21/8}}(r^5+r_c^5)^{1/4}
\end{align}
where $r_0$ is the radius of the shell and $r$ is the radial distance from the center of the shell. Here, $r_c$ is the cross-over point, which characterizes the region of influence of the D2 shell. Far away from the shell $r \gg r_c$, the metric \eqref{e3.18} approaches the near horizon geometry of N D0 brane \eqref{e3.1} and does not depend on the individual configurations of D2 shell.

We consider the massive probe following the radial trajectory $r=r(t)$ while falling close to the configuration of the shell of D2 branes. The point particle Lagrangian (in Einstein's frame) can be expressed as
\begin{align}
    &S_p = \int dt L ~;~L=-\sqrt{h_{tt}(r)-h_{rr}(r)\dot{r}^2}\\
    &h_{tt}(r)=\beta_t r^{7/4}_0 r^{25/8}(r^5+r_c^5)^{1/4}~;~h_{rr}(r)=\beta_r \frac{(r^5+r_c^5)^{1/4}}{r_0^{1/4}r^{15/8}}\\
    & \beta_t = \frac{10^{7/8}}{\sqrt{g_s}R^{49/8}}~;~ \beta_r = \frac{R^{7/8}}{\sqrt{g_s}10^{1/8}}.
\end{align}

Next, we note the Hamiltonian constraint, which can be inverted to obtain the trajectory ($r(t)$) of the massive probe near the shell of D2 branes
\begin{align}
    H=\frac{h_{tt}}{\sqrt{h_{tt}-h_{rr}\dot{r}^2}}\Rightarrow \dot{r}(t)=-\sqrt{\frac{h_{tt}}{h_{rr}}}\sqrt{1-\frac{h_{tt}(r)}{h_{tt}(r_{0})}}.
\end{align}

The regime in which we are interested corresponds to $r\ll r_0$ \cite{Lin:2004kw} and as a result we can always ignore the sub-leading corrections under the square root. This yields the following equation for the radial trajectory ($r(t)$) of the particle
\begin{align}
\label{e3.24}
    -\int \frac{dr}{r^{5/2}}=\beta t \Rightarrow r(t)=\Big( \frac{2}{3 \beta}\Big)^{2/3}t^{-2/3}
\end{align}
where $t\gg 1$ and $\beta=\frac{\sqrt{10}r_0}{R^{7/2}}$.

In order to compute the proper radial momentum ($\mathcal{P}_\rho$) in the vicinity of the D2 shell, we first note the proper distance along the geodesic
\begin{align}
\label{e3.25}
    d\rho = \sqrt{h_{rr}}dr \Rightarrow \frac{\dot{\rho}}{\dot{r}}=\sqrt{h_{rr}}.
\end{align}

Using \eqref{e3.25}, the proper momentum can finally be expressed as
\begin{align}
    \frac{d \mathcal{C}}{dt}\Big|_{t \gg 1}=\mathcal{P}_\rho |_{t \gg 1} =P_r \frac{\partial \dot{r}}{\partial \dot{\rho}}=\frac{H}{\sqrt{h_{tt}}}.
\end{align}

Considering an expansion close to the D2 branes $r \sim r_c$ and using the solution \eqref{e3.24}, it is straightforward to show the rate of growth at late times as
\begin{align}
   \frac{d \mathcal{C}}{dt}\Big|_{t \gg 1}= k t^{35/24}~;~k=H \frac{g_s^{1/4}3^{25/24}10^{7/9}r_0^{7/12}}{R^{49/48}2^{19/12}}.
\end{align}

Clearly, the rate of growth of complexity differs from its UV behavior \eqref{e3.17}, which clearly reveals a non-linear growth of complexity in the IR. This is further confirmed by taking a second derivative of the complexity, which reveals that
\begin{align}
   \frac{d^2 \mathcal{C}}{d^2t}\Big|_{t \gg 1}=\frac{35 k}{24}t^{11/24}.
\end{align}

\section{A comment on non-Abelian T-duality and matrix models}
\label{section 4}
As a final example, we focus on the non-Abelian T-dual of $AdS_5 \times S^5$ \cite{Lozano:2017ole} that is dual to the irrelevant deformation of the matrix model and that does not asymptote to the D0 brane and instead refers to the \emph{smeared} D0 brane asymptotic. The corresponding metric and the dilaton in the string frame can be expressed as
\begin{align}
\label{e4.1}
    &ds^2=-4 \sigma^2 dt^2+\frac{4}{\sigma^2 -1}(d\sigma^2 + d\eta^2)+\frac{4\eta^2 (\sigma^2-1)}{4\eta^2+(\sigma^2-1)^2}d\Omega^2_2+4d\Omega^2_5\\
    &e^{\phi}= \frac{2}{\left(\sigma ^2-1\right)^{1/2} \left(4 \eta ^2+\left(\sigma ^2-1\right)^2\right)^{1/2}}.
\end{align}

The next step is to write the metric \eqref{e4.1} in the Einstein frame, and the rest of the discussions go identically to those given in  \eqref{e2.8}-\eqref{e2.20}. The metric components in the Einstein's frame read as
\begin{align}
    h_1(\sigma , \eta)=\frac{2 \sqrt{2}   \sigma ^2}{ \sqrt[8]{\frac{1}{\left(\sigma ^2-1\right)^2 \left(4 \eta ^2+\left(\sigma ^2-1\right)^2\right)^2}}}~;~h_2(\sigma , \eta)=\frac{2 \sqrt{2} }{\left(\sigma ^2-1\right) \sqrt[8]{\frac{1}{\left(\sigma ^2-1\right)^2 \left(4 \eta ^2+\left(\sigma ^2-1\right)^2\right)^2}}}.
\end{align}

Using the Hamiltonian constraint \eqref{e2.15}, we rewrite \eqref{e2.16}-\eqref{e2.17} as
\begin{align}
\label{e4.4}
    \frac{d}{dt}\Big[ \frac{h_2\dot{\sigma}}{h_1}\Big]+\frac{1}{2h_1}\Big[ \partial_\sigma h_1 -\partial_\sigma h_2 (\dot{\sigma}^2+\dot{\eta}^2)\Big]=0\\
    \frac{d}{dt}\Big[ \frac{h_2\dot{\eta}}{h_1}\Big]+\frac{1}{2h_1}\Big[ \partial_\eta h_1 -\partial_\eta h_2 (\dot{\sigma}^2+\dot{\eta}^2)\Big]=0.
    \label{e4.5}
\end{align}

One has to solve the above set of equations for large and small values of the coordinates and compute the corresponding rates of complexities/proper momentum. This will yield the late time and early time growth of complexity. Expanding the metric components and their derivatives about $\eta = 0$ axis, one finds
\begin{align}
   & h_2(\eta \sim 0, \sigma)=\frac{2 \sqrt{2}}{(\sigma^2 -1)^{1/4}}~;~ \frac{\partial_\sigma h_2}{2 h_1}|_{\eta \sim 0}= -\frac{1}{4 \sigma  \left(\sigma ^2-1\right)^2}\\
   & h_1(\eta \sim 0, \sigma)= 2 \sqrt{2}\sigma^2 \left(\sigma^2 -1\right)^{3/4}~;~\frac{\partial_\sigma h_1}{2h_1}|_{\eta \sim 0}=\frac{7 \sigma ^2-4}{4 \sigma ( \sigma ^2-1)}\\
   & \frac{h_2}{h_1}|_{\eta \sim 0}= \frac{1}{\sigma ^2 (\sigma^2 -1)}~;~\frac{\partial_\eta h_2}{2h_1}|_{\eta \sim 0}=0~;~\frac{\partial_\eta h_1}{2h_1}|_{\eta \sim 0}=0.
\end{align}

We want to find a consistent solution to the above equations of motion \eqref{e4.4}-\eqref{e4.5} subjected to the condition $\dot{\eta}=\ddot{\eta}=0$. In other words, in our calculation, the particle starts in UV ($\sigma =\sigma_{UV} \rightarrow \infty$) and thereby moves towards the center ($\sigma=1$). Imposing this condition in \eqref{e4.5}, one finds that the $\eta$- equation of motion is trivially satisfied.

\begin{figure}
    \centering
    \includegraphics[width=0.5\linewidth]{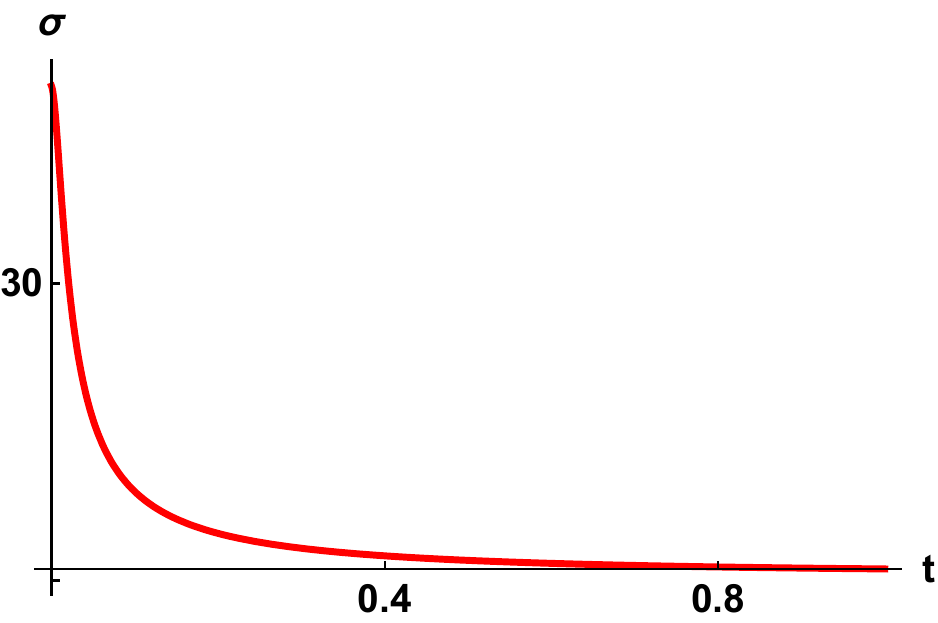}
    \caption{ $\sigma$ vs. $t$ plot. Here we set $\sigma_{UV}=50$.}
    \label{figsigma}
\end{figure}

\begin{figure}
    \centering
    \includegraphics[width=0.5\linewidth]{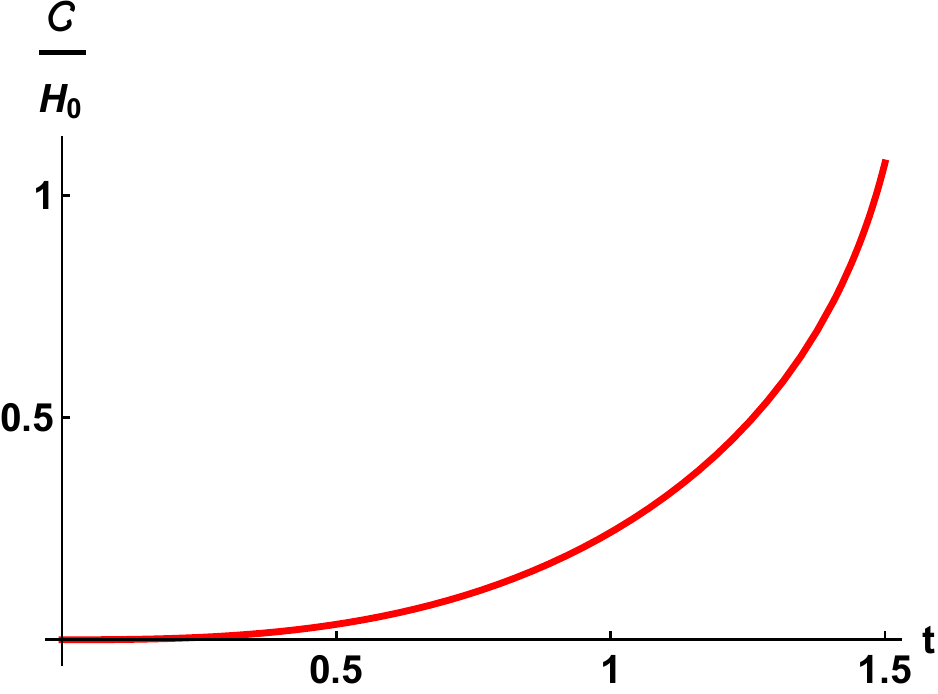}
    \caption{ $\mathcal{C}(t)$ vs. $t$ plot. Here we set $\sigma_{UV}=50$.}
    \label{figCnatd}
\end{figure}

In other words, $\eta = \dot{\eta}=\ddot{\eta}=0$ is a solution to the $\eta$- equation of motion \eqref{e4.5}. Imposing this at the level of the Hamiltonian constraint \eqref{e2.15}, one finds a first order differential equation for $\sigma (t)$, which is equivalent to \eqref{e4.4}
\begin{align}
\label{e4.9}
    \frac{d \sigma}{dt}=-\sigma \sqrt{\sigma^2 -1}\sqrt{1-\frac{\sigma^2}{\sigma^{7/2}_{UV}}(\sigma^2 -1)^{3/4}}.
\end{align}

The solution corresponding to \eqref{e4.9} is shown in Fig.\ref{figsigma}. As Fig.\ref{figsigma} reveals, the particle starts near $\sigma_{UV}$, which falls at later instant of time towards the center of the bulk ($\sigma \sim 1$). Like before, the above solution is subject to the fact that the initial velocity of the probe particle along the $\sigma$ axis is zero ($\dot{\sigma}(t=0)=0$).

Next, we compute the proper momentum \eqref{e2.20} (with $\dot{\eta}=0$), together with \eqref{e2.15}, which finally yields an integral of the form
\begin{align}
    \mathcal{C}(t)/H_0 = \int \frac{dt}{\sqrt{h_1}}\sqrt{1-\frac{\sigma^2}{\sigma^{7/2}_{UV}}(\sigma^2 -1)^{3/4}}.
\end{align}

The corresponding complexity plot is shown in Fig.\ref{figCnatd}. The complexity grows slowly at initial time, while at late time it shoots as the particle approaches the singularity near $\sigma \sim 1$. This behavior differs significantly from the previous examples. 
\section{Krylov complexity for BMN matrix model}
\label{section 5}
Before we conclude, it is important to outline a bridge between the gravity calculations and the matrix model counterpart. In particular, it is important to understand whether the bulk calculations have a manifestation in terms of a Krylov quantum chain for the matrix model counterpart. If so, then how could this be realized in a simple set up.

Given the present state of the art, here we outline a calculation that can be pursued for the matrix model counterpart. The Krylov growth of the Operator ($\mathcal{O}(t)$) in the matrix model is presumably on a basis related to the Krylov complexity, where the spread of the operator could be thought of as an expansion of the form $|\mathcal{O}(t))=\sum_n i^n \phi_n(t)|K_n)$, where the Krylov basis $|K_n)$ satisfies \eqref{ee5.34}, such that it tri-diagonalizes the \emph{Liouvillian} and boils down into the Krylov chain condition for the coefficients $\phi_n$ that satisfy the Schrodinger equation \cite{Parker:2018yvk}. Here, $\hat{\mathcal{L}}=[\hat{H},]$ is the Liouvillian super operator such that $\hat{\mathcal{L}}|K)=|[\hat{H},K])$.

The bosonic part of the Lagrangian can be expressed as \cite{Asano:2015eha}-\cite{Amore:2024ihm} 
\begin{align}
\label{e5.1}
    S=\int dt \text{Tr}\Big[\frac{1}{2}(D_t X^A)^2+\frac{1}{4}[X^A,X^B]^2-\frac{1}{2}\Big( \frac{\mu}{3}\Big)^2 X^2_i -\frac{1}{2}\Big( \frac{\mu}{6}\Big)^2 X^2_a -i \frac{\mu}{3}\epsilon_{ijk}X^i X^j X^k\Big]
\end{align}
where $A,B=1, \cdots 9$. Here, $X_a (a=4,\cdots ,9)$ are the $SO(6)$ scalars and $X_i (i=1,2,3)$ are the $SO(3)$ scalars. The operator $\mathcal{O}$, is $U(N)$ invariant, which transforms like $\mathcal{O}\rightarrow U^{-1}\mathcal{O}U$. 
\subsection{Pulsating fuzzy sphere model}
The action \eqref{e5.1} represents a massive deformation of the BFSS matrix model \cite{Banks:1996vh}, where $\mu$ stands for the mass parameter. The theory has a trivial vacuum ($X^A=0$) as well as \emph{fuzzy} spheres as a solution to the classical equations of motion. Here, $D_t=\partial_t-i[A,]$ is the gauge covariant derivative. With $A=0$, the classical configuration can be boiled down into simple systems, for example, pulsating fuzzy spheres, characterized by the ansatz \cite{Asano:2015eha}
\begin{align}
\label{e5.2}
    &X^i=y(t)\frac{\sigma^i}{2}~;X^{a'}=x(t)\frac{\sigma^{a'-3}}{3}, (a'=4,5,6)\\
    &X^7=X^8=X^9=0.
    \label{e5.3}
\end{align}
The above configuration \eqref{e5.2}-\eqref{e5.3} corresponds to a system of two coupled non-linear oscillators and is a simple example of the $N=2$ matrix model. 

The system is characterized by the Lagrangian density of the following form \cite{Amore:2024ihm}
\begin{align}
    L=\frac{\dot{x}^2}{2}+\frac{\dot{y}^2}{2}-\frac{\mu^2}{8}x^2-\frac{\mu^2}{2}y^2-\frac{1}{2}(x^4+y^4)-x^2 y^2 +\mu y^3.
\end{align}

The corresponding Hamiltonian is given by
\begin{align}
    &H=\frac{p^2_x}{2}+\frac{p_y^2}{2}+V(x,y)\\
    &V(x,y)=\frac{\mu^2}{8}x^2+\frac{\mu^2}{2}y^2+\frac{1}{2}(x^4+y^4)+x^2 y^2 -\mu y^3.
\end{align}

Clearly, in the operator (or Heisenberg) formalism, one has to promote $(x ,p) \rightarrow (\hat{x},\hat{p})$, which leads to the Hamiltonian operator $H \rightarrow \hat{H}$. Therefore, we can define a corresponding Liouvillian operator as $\hat{\mathcal{L}}=[\hat{H},]$. The non trivial task is, however, to find an operator $\mathcal{O}\rightarrow K$ and hence a basis $|K_n)$ such that $(K_n|\hat{\mathcal{L}}|K_n)=0$ and is non-zero for $m\neq n$.
\subsection{Constructing the Krylov basis}
As a trail, we consider the normalized Gaussian operator as our initial state\footnote{We are working in a unit $c=\hbar =l=1$.} \cite{Hashimoto:2023swv}
\begin{align}
\label{e5.7}
    \mathcal{O}_0=\sqrt{\frac{2}{\pi \alpha}}e^{-\frac{\hat{x}^2+\hat{y}^2}{\alpha}}=\frac{1}{\sqrt{N_0}}e^{-\frac{\hat{x}^2+\hat{y}^2}{\alpha}}.
\end{align}
Notice that the above ansatz \eqref{e5.7} is valid for $\mu \neq 0$, where the constant $\alpha^{-1} =\mu$ should be identified with the mass of the oscillator(s).

Given two operators $\mathcal{O}_n$ and $\mathcal{O}_m$, we define the trace as the following integral 
\begin{align}
    \text{Tr}(\mathcal{O}_n \mathcal{O}_m)=\int dx dy \bra{x,y}\mathcal{O}_n \mathcal{O}_m \ket{x,y}
\end{align}
which is taken over the position eigen states in the Hilbert space \cite{Hashimoto:2023swv}.

The Gaussian operator \eqref{e5.7} is normalized to unity. This can be checked following the definition of the inner product in the Hilbert space of operators \cite{Parker:2018yvk},\cite{Hashimoto:2023swv}
\begin{align}
    (\mathcal{O}_0|\mathcal{O}_0)=\text{Tr}(\mathcal{O}_0^\dagger \mathcal{O}_0)=\frac{2}{\pi \alpha}\int dx dy ~e^{-\frac{2}{\alpha}(x^2+y^2)}=1.
\end{align}

In what follows, given the normalized operator \eqref{e5.7}, we obtain the first few operators $\mathcal{O}_n (n=1,2)$ by sequentially applying the Liouvillian operator $\hat{\mathcal{L}}=[\hat{H},]$, where we assume the prescription to be valid for any generic operator $|\mathcal{O}_n)=\hat{\mathcal{L}}^n|\mathcal{O}_0)$. As a result of this procedure, we generate a basis $\{ |\mathcal{O}_n)\}(n=0,1,2,\cdots)$ in the Hilbert space of operators, which does not satisfy the orthogonality of the Krylov basis, that is, $L_{nm}=(\mathcal{O}_n|\hat{L}|\mathcal{O}_m)= 0$ for $n=m$ \cite{Parker:2018yvk}. This can be verified using the following procedure. 

The operators $\mathcal{O}_n$ can be constructed taking the commutator with \eqref{e5.7}
\begin{align}
    \mathcal{O}_1&=[\hat{H},\mathcal{O}_0]=\frac{2\sqrt{2} i}{\sqrt{\pi} \alpha^{3/2}}\Big (\hat{p}_x \hat{x} + \hat{p}_y \hat{y}\Big)e^{-\frac{\hat{x}^2+\hat{y}^2}{\alpha}}\\
    \mathcal{O}_2&=[\hat{H},\mathcal{O}_1]=\frac{2\sqrt{2} }{\sqrt{\pi} \alpha^{3/2}}\Big[ \hat{p}^2_x +\hat{p}^2_y -\frac{2}{\alpha}(\hat{p}_x \hat{x}+\hat{p}_y \hat{y})^2\Big]e^{-\frac{\hat{x}^2+\hat{y}^2}{\alpha}}-\frac{\mu^2}{ \sqrt{2 \pi} \alpha^{3/2}}(\hat{x}^2+4 \hat{y}^2)e^{-\frac{\hat{x}^2+\hat{y}^2}{\alpha}}\nonumber\\
    &-\frac{4\sqrt{2}}{\sqrt{\pi} \alpha^{3/2}}(\hat{x}^4 + \hat{y}^4)e^{-\frac{\hat{x}^2+\hat{y}^2}{\alpha}}-\frac{8\sqrt{2}}{\sqrt{\pi} \alpha^{3/2}}\hat{x}^2 \hat{y}^2e^{-\frac{\hat{x}^2+\hat{y}^2}{\alpha}}+\frac{6\sqrt{2} \mu}{\sqrt{\pi} \alpha^{3/2}}\hat{y}^3e^{-\frac{\hat{x}^2+\hat{y}^2}{\alpha}}
\end{align}
and so on. Here we use $[\hat{x},\hat{p}_x]=i=[\hat{y},\hat{p}_y]$ with $\hat{p}_i=-i \partial_i$ together with $\hbar=1$.

As a trivial check, one can see that the basis $\{|\mathcal{O}_n)\}$ is not orthogonal, therefore, it does not satisfy the Krylov basis criteria \cite{Parker:2018yvk}. For example, one can show the following
\begin{align}
   L_{00}= (\mathcal{O}_0|\hat{\mathcal{L}}|\mathcal{O}_0)=\text{Tr}(\mathcal{O}_0^\dagger \mathcal{O}_1)=2 \mu.
\end{align}

As a further check, the next diagonal element $L_{11}$ can be computed. The massive contribution to $L_{11}$ can be expressed separately as
\begin{align}
\label{e5.12}
   (\mathcal{O}_1|\hat{\mathcal{L}}|\mathcal{O}_1)|_{\mu}&=\text{Tr}(\mathcal{O}_1^\dagger \mathcal{O}_2)|_{\mu}\nonumber\\
   &= \frac{48 \mu}{\pi \alpha^3}\int dx dy \Big[y^3-\frac{\mu}{12}(x^2+4 y^2) \Big]\Big[1-\frac{x^2 +y^2}{\alpha} \Big] e^{-\frac{2}{\alpha}(x^2+y^2)}.
\end{align}

A straightforward evaluation of the integral \eqref{e5.12} yields
\begin{align}
    (\mathcal{O}_1|\hat{\mathcal{L}}|\mathcal{O}_1)|_{\mu}=0
\end{align}
which suggests that corrections due to massive deformation vanishes identically.

One can proceed to compute other contributions in the diagonal element, which finally reveals the $L_{11}$ matrix element of the Liouville operator ($\hat{\mathcal{L}}$) for non-orthogonal state $\mathcal{O}_1$
\begin{align}
  L_{11}=  (\mathcal{O}_1|\hat{\mathcal{L}}|\mathcal{O}_1)&=\text{Tr}(\mathcal{O}_1^\dagger \mathcal{O}_2)=4(1+8 \mu^3)
\end{align}
which clearly suggests a change of basis $|\mathcal{O}) \rightarrow |K)$, such that $L_{11}=0$.\\\\
\uline{\textbf{Gram-Schmidt orthogonalization}}\\\\
In order to define a Krylov basis, one has to construct a proper orthogonal set of states $\{|K_n)\}$ using previously introduced states $\{|\mathcal{O}_n)\}$, which ensures that the diagonal entries of the Liouvillian operator ($\hat{\mathcal{L}}$) are zero \cite{Parker:2018yvk}. This is achieved following the Gram-Schmidt orthogonality procedure. We define the following linear map (for $n \geq 0$)
\begin{align}
\label{e5.15}
    | K_{n+1})=|\mathcal{O}_{n+1})-c_n | K_{n})-d_{n}|K_{n-1}).
\end{align}

The first few Krylov states (for $n=0,1$) can be expressed as
\begin{align}
\label{e5.16}
    &|K_1)=|\mathcal{O}_1)-c_0 |K_0)\\
    &|K_2)=|\mathcal{O}_2)-c_1|K_1)-d_1 |K_0)
    \label{e5.17}
\end{align}
which is subject to the fact $|{K_0})=|\mathcal{O}_0)$.

Taking the inner product with $|K_0)$ and setting $(K_0|K_1)=0$, we obtain the coefficient
\begin{align}
    c_0 = \text{Tr}(\mathcal{O}^\dagger_0 \mathcal{O}_1)=2 \mu.
\end{align}

On the other hand, from \eqref{e5.17}, taking the inner product with $|K_0)$ we obtain
\begin{align}
    (K_0|K_2)=0=(K_0|\mathcal{O}_2)-d_1
\end{align}
which implies that the above coefficient can be expressed as
\begin{align}
    d_1=\text{Tr}(\mathcal{O}^\dagger_0 \mathcal{O}_2)=\frac{27}{8}\mu^2 -\frac{2}{\mu}.
\end{align}

Finally, taking the inner product with $|K_1)$ we obtain
\begin{align}
   (K_1|K_2)=0=(K_1|\mathcal{O}_2)-c_1
\end{align}
which yields the following coefficient
\begin{align}
    c_1=\text{Tr}(\mathcal{O}^\dagger_1 \mathcal{O}_2)-c_0 \text{Tr}(\mathcal{O}^\dagger_0 \mathcal{O}_2)=8+\frac{101}{4}\mu^3
\end{align}
where we have used the orthonormal nature of $|K_1)$, that is, $(K_1|K_1)=1$.

Clearly, the disappearance of the first diagonal entry $L_{11}$ in the new (Krylov) basis follows from the orthogonality of the states $|K_1)$ and $|K_2)$
\begin{align}
\label{e5.18}
    L_{11}=(K_1|\hat{\mathcal{L}}|K_1)=(K_1|K_2)=0.
\end{align}

We normalize the state by rescaling $|K_1)\rightarrow \frac{1}{\sqrt{N_1}}|K_1)$, so that $(K_1|K_1)=1$. A straightforward calculation reveals the norm of the state
\begin{align}
\label{e5.25}
    (K_1|K_1)=N_1 = 4\mu^2.
\end{align}

Next, we have to introduce a normalized basis state as $|K_2)\rightarrow\frac{1}{\sqrt{N_2}}|K_2)$ such that $ (K_2|K_2)=1$. A straightforward calculation reveals the following
\begin{align}
    (K_2|K_2)=N_2=\text{Tr}(\mathcal{O}^\dagger_2 \mathcal{O}_2)+c^2_1\Big(1-\frac{1}{\mu}\Big) -d^2_1.
\end{align}

Individual entities may be evaluated separately, which yields
\begin{align}
\label{e5.27}
    N_2&=\frac{59 \mu ^4}{64}+\left(\frac{101 \mu ^3}{4}+8\right)^2 \left(1-\frac{1}{\mu }\right)-\left(\frac{27 \mu ^2}{8}-\frac{2}{\mu }\right)^2\nonumber\\
    &+\frac{5}{16} \left(\frac{51}{\mu ^3}+32\right) \mu ^4+8 \left(\frac{3}{\mu ^6}+\frac{6}{\mu ^3}+28\right) \mu ^4.
\end{align}

In summary, we have orthonormal (Krylov) states $|K_n)$ for $n=0,1,2$ 
\begin{align}
    & |K_0)=| \mathcal{O}_0)\\
    \label{e5.29}
    &|K_1)=\frac{1}{\sqrt{N_1}}\Big[| \mathcal{O}_1)-2 \mu |K_0)\Big]\\
    &|K_2)=\frac{1}{\sqrt{N_2}}\Big[|\mathcal{O}_2)-\Big(8+\frac{101}{4}\mu^3\Big)|K_1)-\Big( \frac{27}{8}\mu^2 -\frac{2}{\mu}\Big) |K_0)\Big]
    \label{e5.30}
\end{align}
so that $(K_n|K_m)=\delta_{nm}$ and $L_{mn}=0$ for $m=n$, which produce the Lanczos coefficient(s) $b_n$ for $m\neq n$. The above procedure can be extended for other values of $n \geq 3$. However, this would be a more challenging task as far as analytic techniques are concerned, which we therefore postpone for future investigation. 
\subsection{Lanczos coefficients and Krylov complexity}
In our computation, we will be mostly concerned with the early time ($t \sim 0$) growth of the Krylov complexity ($\mathcal{C}(t)$). Therefore, it is sufficient for us to compute the first few Lanczos coefficients ($b_n$) in the expansion. In the following, we estimate them for $n=0,1,2$.

Given the Krylov basis $\{ | K_n)\}$, the diagonal entries are all zero
\begin{align}
    a_n=L_{nn}=(K_n|\hat{\mathcal{L}}|K_n)
\end{align}
where we identify that the following diagonal elements are zero by construction
\begin{align}
    &a_0=(K_0|\hat{\mathcal{L}}|K_0)=(K_0|K_1)=0\\
    &a_1=(K_1|\hat{\mathcal{L}}|K_1)=(K_1|K_2)=0.
\end{align}

The other set of Lanczos coefficients ($b_n$) are fixed by the Krylov chain condition \cite{Parker:2018yvk}
\begin{align}
\label{ee5.34}
    &|A_{n+1})=\hat{\mathcal{L}}|K_n)-b_{n}|K_{n-1})\\
    &|A_{n+1})=b_{n+1}|K_{n+1}).
\end{align}

 Taking the inner product with $|K_{n-1})$, they are given by the off-diagonal entries \cite{Hashimoto:2023swv}
\begin{align}
    b_n=L_{n n-1} = (K_n| \hat{\mathcal{L}}|K_{n-1})
\end{align}
where we have also used the cyclic property of trace.

Clearly, for $n=0$ the state $|K_{-1})=0$ and we have $b_0=0$. On the other hand, for $n=1$, we obtain the first non-zero Lanczos coefficient (with $K_0=\mathcal{O}_0$)
\begin{align}
    b_1=L_{10}=(K_1|\hat{\mathcal{L}}|K_0)=(K_1|[H,K_0])=(K_1|\mathcal{O}_1).
\end{align}

Using \eqref{e5.29}, we can further simplify the above expression as
\begin{align}
    b_1=\frac{1}{\sqrt{N}_1}\Big[(\mathcal{O}_1|\mathcal{O}_1)-2\mu (\mathcal{O}_0|\mathcal{O}_1)\Big]=2\mu.
\end{align}

Notice that $b_1$ is a linear function and purely fixed by the mass deformation parameter $\mu$. The second non-vanishing Lanczos coefficient corresponds to $n=2$ 
\begin{align}
    b_2 =L_{21}=(K_2|\hat{\mathcal{L}}|K_1)=(K_2|[H,K_1]).
\end{align}

Using \eqref{e5.29}, this further yields
\begin{align}
    b_2&=\frac{1}{\sqrt{N}_1}(K_2|[H,\mathcal{O}_1])-\frac{2\mu}{\sqrt{N_1}}(K_2|[H,K_0])\nonumber\\
    &=\frac{1}{2\mu}(K_2|\mathcal{O}_2)-(K_2|\mathcal{O}_1).
\end{align}

Using \eqref{e5.30} we finally obtain the following expression
\begin{align}
    b_2(\mu)=\frac{1}{64 \mu ^3 \sqrt{N_2}}\Big[ -6969 \mu ^7+5537 \mu ^6-5440 \mu ^4+1966 \mu ^3-1024 \mu +640\Big].
\end{align}

\begin{figure}
    \centering
    \includegraphics[width=0.5\linewidth]{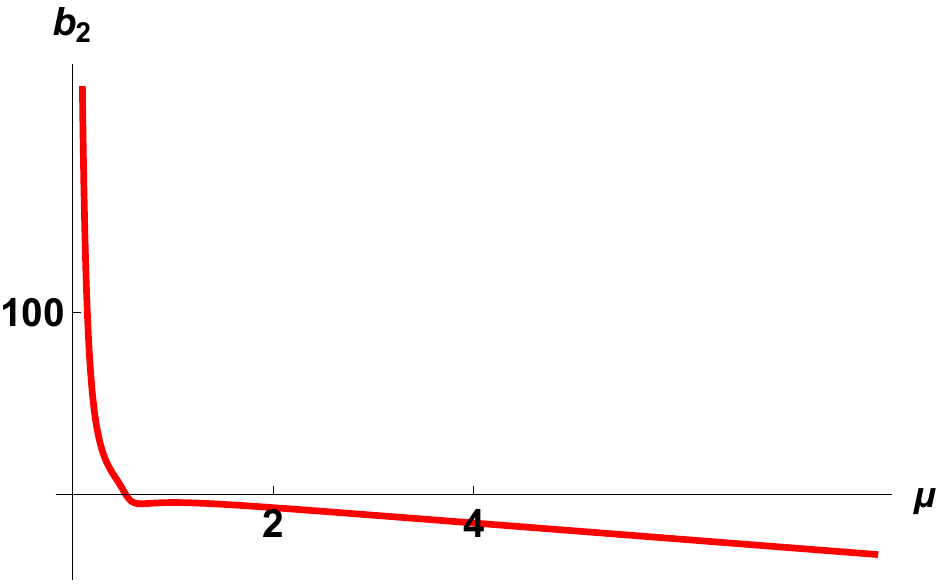}
    \caption{We plot the second Lanczos coefficient $b_2$ with the mass parameter $\mu$.}
    \label{figb2}
\end{figure}

Clearly, the coefficient $b_2$ scales differently in the small and large $\mu$ limits. In the limit $\mu \ll 1$, it scales as $b_2= \frac{\sqrt{5}}{\mu^2} $. The coefficient $b_2$ decreases with the increase in $\mu$ and eventually there is a crossover from $b_2>0$ to $b_2<0$ at some critical $\mu=\mu_c$ (see Fig.\ref{figb2}). For large $\mu \gg 1$, the coefficient $b_2$ scales linearly with $\mu$, that is, $b_2=-\frac{69}{16}  \mu +\text{constant}$ . 

Next, we solve the coefficients $\phi_n(t)$, which satisfy the Schrodinger equation \cite{Parker:2018yvk} 
\begin{align}
    \partial_t \phi_n =b_n \phi_{n-1}(t) -b_{n+1}\phi_{n+1}(t).
\end{align}

Taking into account $n=0,1,2$, one arrives at the following set of equations
\begin{align}
\label{e5.32}
    & \partial_t \phi_0 = -b_1 \phi_1 (t)\\
    &\partial_t \phi_1 =b_1 \phi_0 (t) -b_2 \phi_2 (t)\\
    & \partial_t \phi_2 = b_2 \phi_1 (t)-b_3 \phi_3 (t).
    \label{e5.34}
\end{align}

The above set of equations \eqref{e5.32}-\eqref{e5.34} can be combined to obtain
\begin{align}
    (\partial^2_t +b^2)\phi_1 (t)= \hat{b}^2 \phi_3(t)
\end{align}
where $b^2(\mu)=b_1^2 + b_2^2 (\mu)$ and $\hat{b}^2 (\mu)=b_2 b_3$.

The corresponding solution can be obtained in terms of the Green's function 
\begin{align}
\label{e5.36}
    \phi_1(t)=-\hat{b}^2\int d t_0 \phi_3(t_0)G(t,t_0).
\end{align}

The Green's function satisfies the inhomogeneous equation of the form
\begin{align}
\label{e5.37}
    (\partial^2_t +b^2)G(t,t_0)=-\delta (t-t_0).
\end{align}

The Green's function satisfies the Neumann boundary condition, namely
\begin{align}
    G(t, t_0)|_{t=0}=\text{constant} ~;~\partial_t G (t, t_0)|_{t=0}=0.
\end{align}

Moving to the frequency ($\omega$) space, we notice that
\begin{align}
\label{e5.39}
    G(t, t_0)=\frac{1}{2\pi}\int d \omega e^{i \omega (t-t_0)}G(\omega) ~;~ \delta (t-t_0)=\frac{1}{2\pi}\int d\omega e^{i \omega (t-t_0)}.
\end{align}

Using \eqref{e5.37} and \eqref{e5.39}, one finds 
\begin{align}
    G(\omega)=\frac{1}{\omega^2 -b^2}.
\end{align}

Taking into account an early time expansion $t  \sim 0$, we can rewrite \eqref{e5.36} as
\begin{align}
\label{e5.41}
    \phi_1 (t \sim 0)=-\hat{b}^2 \int_0^t dt \phi_3 (t)G(0,t).
\end{align}

Expanding the arguments in the integral \eqref{e5.41} for $t=0$, one finds at leading order
\begin{align}
   \phi_1 (t \sim 0)=-\hat{b}^2\int_0^t dt \phi_3(0)G(0,0)+\cdots . 
\end{align}

A careful analysis reveals the following
\begin{align}
\label{e5.43}
    G(0,0)=\frac{1}{2 \pi}\int d\omega \frac{e^{-i \omega t}}{\omega^2 -b^2}\Big|_{t = 0}=-\frac{\tanh ^{-1}\left(\frac{\omega }{b}\right)}{2 \pi  b}.
\end{align}

Using \eqref{e5.43}, the leading contribution to the Krylov complexity appears to be
\begin{align}
\label{e5.44}
    \mathcal{C}(t)|_{t \sim 0}=|\phi_1 (t \sim 0)|^2+ \cdots = c_1 t^2 + \cdots
\end{align}
where the leading term exhibits a quadratic growth, identical to \eqref{e2.43}.

The coefficient of the leading term can be expressed in terms of Lanczos coefficients
\begin{align}
\label{e5.47}
    c_1 (\mu) = \frac{b^2_2 b^2_3}{4 \pi^2 (b^2_1 +b_2^2)}|\phi_3(0)|^2 \Big|\tanh ^{-1}\left(\frac{\omega }{b}\right) \Big|^2
\end{align}
where each of the Lanczos coefficients $b_i$ above depends on the mass parameter $\mu$. Notice that the leading coefficient \eqref{e5.47} depends on the mass parameter $\mu$, which on the gravity side is reflected in the dipole deformation $P$, see for example eq. \eqref{e2.43}.

Clearly, the Krylov complexity \eqref{e5.44} is corrected due to massive deformation ($\mu$). The Lanczos coefficient $b_3$, is given by the following expression
\begin{align}
    b_3 = L_{32}=(K_3| \hat{\mathcal{L}}|K_{2}).
\end{align}

The Krylov basis element $|K_3)$ is fixed by the orthonormality condition, which satisfies the Krylov chain \eqref{e5.34}. Following \eqref{e5.15}, we express the Krylov basis element for $n=3$
\begin{align}
    |K_3)=|\mathcal{O}_3)-c_2 |K_2)-d_2 |K_1)
\end{align}
where $|\mathcal{O}_3) = |[\hat{H},\mathcal{O}_2)]$. The constants $c_2$ and $d_2$ are fixed by the orthogonality criteria.

Before we conclude, it is worth mentioning some important points that remain to be explored. It would be nice to find a generic algorithm \cite{Hashimoto:2023swv} that determines the Lanczos coefficients $b_n$ for arbitrary $n$ and, in particular, to explore the behavior in the limit $n \rightarrow\infty$. It would be nice to see whether these coefficients scale linearly with $n$, which is a typical characteristic of chaotic systems \cite{Parker:2018yvk}. For the BMN matrix model, similar features should be expected, since the model exhibits chaos \cite{Asano:2015eha}-\cite{Amore:2024ihm}.

It would be nice to explore the behavior of the Lanczos coefficients $b_n$ in the limit of large deformation $\mu \gg 1$. This limit is particularly interesting because the system transits into an integrable domain \cite{Amore:2024ihm}. As our analysis reveals, both $b_1$ and $b_2$ scale linearly with $\mu$ in the domain of large mass deformation. It would be nice to explore whether this is an universal feature and has any characteristic role in classifying the underlying integrable or non-integrable feature of the (fuzzy sphere) matrix model.

It would be nice to extend the above calculations for the full growth of Krylov complexity and in particular to study the signature of chaos following the lines of \cite{Asano:2015eha}-\cite{Amore:2024ihm}. On top of it, the late time growth should show up some resemblance with the predictions from the gravity calculations. The next step would be to improve the algorithm for the full supersymmetric parent theory. We hope to address some of these issues in the near future.

\paragraph{Acknowledgements.}
 The author thanks Carlos Nunez for discussion. The author also acknowledges the Mathematical Research Impact Centric Support (MATRICS) grant no. (MTR/2023/000005) received from ANRF, India. \\ 

\end{document}